\documentclass[12pt]{iopart}

\newcommand{\bea}{\begin{eqnarray}}    
\newcommand{\eea}{\end{eqnarray}}      
\newcommand{\be}{\begin{equation}}
\newcommand{\ee}{\end{equation}}
\newcommand{\bef}{\begin{figure}}
\newcommand{\eef}{\end{figure}}

\usepackage{epsfig} 

\begin{document}

\title{Inhomogeneities in the universe}

\author{Francesco Sylos Labini} 
\address{Centro Enrico Fermi, Piazza
  del Viminale 1,  00184 Rome, Italy and \\ Istituto dei Sistemi
  Complessi CNR, Via dei Taurini 19, 00185 Rome, Italy}
\ead{sylos@centrofermi.it}

\begin{abstract}
{ Standard models of galaxy formation predict that matter
  distribution is statistically homogeneous and isotropic and
  characterized by (i) spatial homogeneity for $r<10$ Mpc/h, (ii)
  small-amplitude structures of relatively limited size (i.e.,
  $r<100$) Mpc/h and (iii) anti-correlations for $r>r_c\approx 150$
  Mpc/h (i.e., no structures of size larger than $r_c$).  Whether or
  not the observed galaxy distribution is interpreted to be compatible
  with these predictions depend on the a-priori assumptions encoded in
  the statistical methods employed to characterize the data and on the
  a-posteriori hypotheses made to interpret the results.  We present
  strategies to test the most common assumptions and we find evidences
  that, in the available samples, galaxy distribution is spatially
  inhomogeneous for $r<100$ Mpc/h but statistically homogeneous and
  isotropic.  We conclude that the observed inhomogeneities pose a
  fundamental challenge to the standard picture of cosmology but they
  also represent an important opportunity which may open new
  directions for many cosmological puzzles.}
\end{abstract}


\maketitle

\section{Introduction}

One of cornerstone of modern cosmology is { represented} by the
observations of the three dimensional distribution of galaxies
\cite{york,colless01}. In recent years { the extraordinary increase
  of the number of redshifts } has allowed { us to characterize in
  detail galaxy structures at low redshifts} (i.e., $z<0.3$) and small
scales (i.e., $r<150$ Mpc/h).  Many authors (see e.g.,
\cite{eisenstein,kazin,cg2011,zehavietal02,zehavi,norbergxi01,norbergxi02})
have { concluded that the results of a statistical analysis of the
  data} are {\it compatible} with the theoretical expectations { of
  standard scenarios of galaxy as the Cold Dark Matter (CDM) model and
  its variants (i.e., the case in which the cosmological constant is
  non zero or LCDM)}.  { However, there are some important
  methodological issues which have not received the due attention
  \cite{sdss_2007,sdss_aea,sdss_epl,2df_aea,copernican,bao_aea}. In
  particular,} the critical points concern the a-priori assumptions
which are usually used, without being directly tested, in the
statistical analysis of the data and the a-posteriori hypotheses that
are invoked to interpret the results.

Among the former, there are the assumptions of spatial homogeneity and
of translational and rotational invariance (i.e., statistical
homogeneity) which are built in the { definition of the} standard
estimators of galaxy correlations \cite{book}. While these estimators
are certainly the correct ones to use when statistical and spatial
homogeneity are verified, { it is not simply evident that
  galaxy data do satisfy these properties in the available samples.}
It is indeed well known that galaxies are organized into { a
  network of} structures, like clusters, filaments and voids, with
large fluctuations
\cite{Ratcliffe98,busswell03,frith03,frith06,saslaw2010,einasto2011}
and it is not a-priori obvious that spatial or statistical homogeneity
are satisfied in a sample of { arbitrary small size}.

{ The observed galaxy distribution is found to be inhomogeneous at
  small scales while, according to theoretical models it is expected
  to become spatially homogeneous for $r> \lambda_0 \approx 10$ Mpc/h
  (see, e.g., \cite{springel}): this scale can be easily calculated by
  considering how the scale at which fluctuations are order of the
  mean evolves according to linear perturbation theory of a
  self-gravitating fluid\cite{pee80}. 
 The scale $\lambda_0$, a key
  theoretical prediction which must be confronted with the data, is
  usually determined only indirectly by using statistical methods
  which assume a-priori spatial homogeneity. When the given 
finite sample distribution {\it is not} spatially homogeneous 
the results of the analysis are very misleading \cite{book}.
Therefore, in order {\it
    to test directly} whether a distribution is spatially homogeneous
  it is necessary to introduce more general statistical methods than
  the usual ones \cite{book,sdss_aea}.  These methods consider
  explicitly the problem of the stability of finite sample
  determinations: if} a statistical quantity depends on the sample
 size then { it is affected by large fluctuations and/or by
observational 
  systematic effects; in both cases it does not represent a}
meaningful and useful estimator of an ensemble average property. A
critical analysis of finite-sample volume averages is thus necessary
to identify the subtle effects induced by spatial inhomogeneities {
  and to distinguish them from other intervening systematic effects}.

As mentioned above, a second kind of ad-hoc hypotheses 
are often used in the interpretation of the { results of the
  statistical analysis. These are invoked when one finds results which
  are a-priori unexpected and which clearly show that some of the
  basic assumptions encoded in the used statistical methods are not
  verified in the data. Examples are galaxy evolution, luminosity
  bias, or selection effects due to some observational issues. It is
  plausible that some of these may affect the results of a statistical
  analysis; however, in the absence of a quantitative prediction or of
  an independent estimation of these effects, one must use several
  assumptions (e.g., specific functional behavior or arbitrary values
  for a set of parameters, etc.)  \cite{blanton,loveday,kazin} without
  any clean test of their validity.  A different strategy, which, when
  possible, we adopt here, is to develop focussed tests to understand
  whether the quantitative influence of the intervening systematic
  effects are supported by the given data.}

{ Theoretical models predict the matter density field properties
  both in the early and in the late universe. Fluctuations and
  correlations must have very specific properties.  Firstly, the}
Friedmann-Robertson-Walker (FRW) geometry is derived under the
assumption that matter distribution is {\it exactly translational and
  rotational invariant}, i.e. that the matter density is assumed to be
constant in a spatial hyper-surface. On the top of the mean field one
can consider statistically homogeneous and isotropic small-amplitude
fluctuations \cite{weinberg}.  These furnish the seeds of
gravitational clustering which eventually give rise to the structures
we observe in the present universe.  

{ Secondly, the} statistical properties of matter density
fluctuations have to satisfy an important condition in order to be
compatible with the FRW geometry \cite{harrison,zeldovich}.  In its
essence, the condition is that fluctuations in the gravitational
potential induced by density fluctuations do not diverge at large
scales \cite{glass,book,cdm_theo}. This situation requires that the
matter density field fluctuations { must} decay in the fastest
possible way with scale \cite{torquato}. Correspondingly the two-point
correlation function becomes negative at larger scales (i.e., $r>150$
Mpc/h) which implies the absence of larger structures of tiny density
fluctuations.  Are the { observed} large scale structures and
fluctuations compatible with such a scenario ?

This paper is organized as follows. In Sect.\ref{prob} we briefly
review the main properties of both spatially homogeneous and
inhomogeneous stochastic density fields.  { The main features of
  real space correlation properties of standard cosmological density
  fields are presented in Sect.\ref{cosmomodels}.  In the case of a
  finite-sample distribution (Sect.\ref{stat}) the information that
  can be exacted from the data is through a statistical analysis, and
  hence through the computation of volume averages}. We discuss how to
set up a strategy to analyze a point distribution in a finite volume,
stressing the sequence of steps that should be considered in order to
reduce as much as possible the role of a-priori assumptions encoded in
the statistical analysis { and to correctly interpret the meaning
  of the measured volume averages}.  The analysis of the galaxy data
is presented in Sect.\ref{results}. We { show} that galaxy
distribution, at relatively low redshifts (i.e., $z<0.3$) and small
scales (i.e., $r<150$ Mpc/h) is characterized by large density
fluctuations which correspond to large-scale correlations.  We
emphasis that by using the standard statistical tools one reaches a
different conclusion. This occurs because these methods are based on
several important assumptions: some of them, when directly tested are
not verified, while others are very strong ad-hoc hypotheses which
require a detailed investigation.  Finally in Sect.\ref{conclusion} we
draw our main conclusions.


\section{A brief review of the main statistical properties} 

\label{prob} 
Before entering in the problems related to the statistical
characterization in finite samples,  we review the main probabilistic
properties of mass density fields. This means that we consider
ensemble averages or, for ergodic cases, volume averages in the
infinite volume limit.

A mass density field can be represented as a stationary stochastic
process that consists in extracting the value of the microscopic
density function $\rho(\vec{r})$\footnote{We use the symbol $\rho(r)$
  for the microscopic mass density and $n(r)$ for the microscopic
  number density. However in the following sections we consider only
  the number density, as it is usually done in studies of galaxy
  distributions. In that case we can simply replace the symbol
  $\rho(r)$ with $n(r)$ and all the definitions given in this section
  remain unchanged.}  at any point of the space. This is completely
characterized by its probability density functional ${\cal
  P}[\rho(\vec{r})]$. This functional can be interpreted as the joint
probability density function (PDF) of the random variables
${\rho}(\vec{r})$ at every point $\vec{r}$.  If the functional ${\cal
  P}[\rho(\vec{r})]$ is invariant under spatial translations then the
stochastic process is {\em statistically homogeneous} or translational
invariant (stationary) \cite{book}.  When ${\cal P}[\rho(\vec{r})]$ is
also invariant under spatial rotation then the density field is {\em
  statistically isotropic} \cite{book}.

%


A crucial assumption usually used, when comparing theoretical
prediction to data, is that stochastic fields are
required to satisfy spatial {\em ergodicity}.  Let us take a generic
observable ${\cal F}= {\cal F}(\rho(\vec{r}_1),\rho(\vec{r}_2),...)$
function of the mass distribution $\rho(\vec{r})$ at different points
in space $\vec{r}_1,\vec{r}_2,...\;$.  Ergodicity implies that 
%
$\left<{\cal F}\right> = \overline{{\cal F}} = \lim_{V
\rightarrow \infty} \overline{{\cal F }}_V \;, 
$
%
where the symbol $\langle...\rangle$ is for the (ensemble) average over 
different realizations of the stochastic process, 
and
%
$\overline{{ \cal F}}_V =  \frac{1}{V} \int_V { \cal F} dV
$
%
is the spatial average in a finite volume $V$ \cite{book}.

\subsection{Spatially homogeneous distributions} 

The condition of {\em spatial homogeneity} ({\em uniformity}) is
satisfied if the ensemble average density of the field $\rho_0=
\langle \rho \rangle$ is strictly positive, i.e. for an ergodic
stochastic field \cite{book}, 
\be 
\langle \rho \rangle = \lim_ {R\rightarrow
  \infty} \frac{1}{V(R;\vec{x_0})} \int_{V(R;\vec{x_0})} \rho(r) d^3r
> 0 \;\; \forall \vec{x_0} \;, 
\ee 
where $R$ is the linear size of a volume $V$ with center in
$\vec{x_0}$.  { Note that it} is necessary to carefully test
spatial homogeneity before applying the definitions given in this
section to a finite sample distribution (see Sect.\ref{stat}).
Indeed, for inhomogeneous distributions the {\it estimation} of the
average density substantially differs from its asymptotic value and
thus the sample estimation of $\rho_0$ is biased by finite size
effects. Unbiased tests of spatial homogeneity can be achieved by
measuring conditional properties (see below).


A distribution is spatially inhomogeneous up to a scale $\lambda_0$,
{ i.e. or spatially homogeneous for $r>\lambda_0$}, if \cite{book}
\be
\label{homoscale} 
\left| \frac{1}{V(R;\vec{x_0})} \int_{V(R;\vec{x_0})} d^3x \rho(\vec{x})
-\rho_0 \right| < \rho_0 \;\; \forall R>\lambda_0 \;\;, \forall
\vec{x_0} \;.  
\ee 
This equation defines the homogeneity scale $\lambda_0$ which
separates the strongly fluctuating regime $r<\lambda_0$ from the
regime where fluctuations have small amplitude relative to the
asymptotic average.

{ Let us now discuss the characterization of two-point correlation
  properties.}  The quantity $\langle
\rho(\vec{r_1})\rho(\vec{r_2})\rangle dV_1 dV_2$ { gives the
  a-priori probability to find} two particles simultaneously placed in
the infinitesimal volumes $dV_1, dV_2$ respectively around $\vec{r_1},
\vec{r_2}$. The quantity \be
\label{cond} 
\langle \rho(r_{12}) \rangle_p dV_1 dV_2 =
\frac{\langle\rho(\vec{r_1})\rho(\vec{r_2})\rangle}{\rho_0} dV_1 dV_2
\ee 
gives the a-priori probability of finding two particles placed in
the infinitesimal volumes $dV_1, dV_2$ around $\vec{r_1}$ and
$\vec{r_2}$ with the condition that the origin of the coordinates is
occupied by a particle (Eq.\ref{cond} is the ratio of unconditional
quantities, and thus, for the roles of { conditional}
probabilities, it defines a conditional quantity) \cite{book}.

For a stationary and spatially homogeneous distribution (i.e.,
$\rho_0>0$),
we may define the reduced two-point correlation function as \cite{book} 
\be 
\label{2point} 
\xi(r_{12}) = \frac{\langle \rho(r_{12}) \rangle_p}{\rho_0} -1   =
\frac{\langle \rho(r_{12}) \rangle}{\rho_0^2} -1  \;.
\ee
This function characterizes { two-point} correlation properties of small
amplitude density fluctuations.  { When spatial homogeneity has
  already been proved there are several useful information that can be
  extracted from $\xi(r)$, and in particular one or a few characteristic 
length scales. For instance, the correlation length
  typically corresponds to an exponential decay of $\xi(r)$ of the
  type $\xi(r) \sim \exp(-r/r_c)$ \cite{book}}.

The two-point correlation function defined by Eq.\ref{2point} is
simply related to the normalized mass variance in a volume $V(R)$ of
linear size $R$ \cite{book}
\be 
\label{eq:sigma2} 
\sigma^2(R) = \frac{\langle M(R)^2\rangle - \langle M(R)
  \rangle^2} {\langle M(R) \rangle^2}  = 
 \frac{1}{V^2(R)} \int_{V(R)}d^3r_1\int_{V(R)} d^3r_2
\xi(r_{12}) \;. 
\ee 
The scale $r_*$ at which
fluctuations are of the order of the mean, i.e.  $\sigma(r_*) =1$, is
proportional to the scale $r_0$ at which $\xi(r_0)=1$ and to the scale
$\lambda_0$ defined in Eq.\ref{homoscale}.

For spatially uniform systems, when the volume $V$ in
Eq.\ref{eq:sigma2} is a real space sphere\footnote{The case in which
  the volume is a Gaussian sphere can be misleading, see { 
    discussion in, e.g., \cite{glass}}}, it is possible to proceed to
the following classification for the scaling behavior of the
normalized mass variance at large enough scales \cite{glass,book}:
\begin{equation}
\sigma^2(R)\sim\left\{
\begin{array}{ll}
R^{-(3+n)} & \mbox{for }\, -3<n<1\\
R^{-(3+1)}\log R & \mbox{for }\, n=1\\
R^{-(3+1)} & \mbox{for }\, n>1
\end{array}
\right. \;. 
\label{sigma5}
\end{equation}

For $-3<n<0$ (which corresponds to $\xi(r) \sim r^{-\gamma}$ with $0 <
\gamma = 3+n < 3$), mass fluctuations are {\em super-Poisson}. { 
  These are, for instance,} typical of systems at the critical point
of a second order phase transition~\cite{book}: there are long-range
correlations and the correlation length $r_c$ is infinite.  For $n=0$
fluctuations are Poisson-like and the system is called {\em
  substantially Poisson}: there are no correlations (i.e., a purely
Poisson distribution) or correlations limited to small scales, { 
  i.e.  of the type $\xi(r) \sim \exp(-r/r_c)$, with a finite
$r_c$}. This behavior is typical of many common physical systems, e.g.,
a homogeneous gas at thermodynamic equilibrium at sufficiently high
temperature. Finally for $n\ge 1$ fluctuations are {\em sub-Poisson}
or {\em super-homogeneous} \cite{glass,book} (or hyper-uniform
\cite{torquato}). In this case $\sigma^2(R)$ presents the fastest
possible decay for discrete or continuous distributions \cite{glass}
and the two-point correlation function has to satisfy { the
following global constraint}

\be
\label{intcos} 
\int_0^{\infty} d^3r \xi(r) = 0 \;,
\ee
{ (see for more details Sect.\ref{cosmomodels})}.  Examples are
provided by the one component plasma, a well-known system in
statistical physics~\cite{lebo}, and by a randomly shuffled lattice of
particles \cite{book,andrea}.

Note that any {\it uniform} stochastic process has to satisfy the
following condition
\be
\label{limitsigma} 
\lim_{R
  \rightarrow \infty} \sigma^2(R) = \lim_{R
  \rightarrow \infty} = 
 \frac{1}{V^2(R)} \int_{V(R)}d^3r_1\int_{V(R)} d^3r_2
\xi(r_{12})
=0
\ee
which implies that the average density $\rho_0$, in the infinite
volume limit, is a well defined concept, i.e. $\rho_0>0$ \cite{book}.
{ This is a weaker condition than that required by Eq.\ref{intcos}.}

\subsection{Spatially inhomogeneous distributions} 

A distribution is spatially inhomogeneous in the ensemble (or in the
infinite volume limit) sense if $\lambda_0 \rightarrow \infty$.  For
statistically homogeneous distributions, from Eq.\ref{homoscale}, { 
  we find } that the ensemble average density is $\rho_0=0$. Thus
unconditional properties are not well defined: if we { consider a
  randomly placed finite volume in an infinite inhomogeneous
  distribution}, it typically contains no points. Therefore only
conditional properties are well defined, as for instance the average
conditional density defined in Eq.\ref{cond}.

For { a statistically homogeneous and isotropic fractal structure
  (where all points are alike)} the average conditional mass included
in a spherical volume grows as $\langle M(r) \rangle_p \sim r^D$: for
$D<3$, the average conditional density presents a scaling behavior of
the type \cite{book}
\be 
\label{scaling} 
\langle \rho(r)
\rangle_p = \frac{\langle M(r) \rangle_p}{V(r)} \sim r^{D-3} \;,
\ee 
so that 
$\lim_{r \rightarrow \infty}  \langle \rho(r)
\rangle_p =0$.  
The hypotheses underlying the derivation of the Central Limit
Theorem are violated by the long-range character of spatial
correlations, resulting in a PDF of fluctuations that does not follow
the Gaussian function \cite{book,gumbel}. On the contrary, { the PDF 
  typically displays ``long tails'' and some moments of the
  distribution may diverge \cite{bouchoud} } .

It is possible to introduce more complex inhomogeneous distributions
than Eq.\ref{scaling}, for instance the multi-fractal distributions
for which the scaling properties are not described by a single
exponent, but they change in different spatial locations \cite{book}.
Another simple (and different !) example is given by a distribution in
which the scaling exponent in Eq.\ref{scaling} depends on distance,
i.e. $D=D(r) < 3$.


\section{Statistical properties of the standard model}  
\label{cosmomodels}

As discussed in the introduction, { the important constraint that must
  be valid for any kind of  matter density fluctuation field in
  the framework of FRW models, is represented} by the condition of
super-homogeneity, corresponding in cosmology to the so-called { 
  property}    of ``scale-invariance'' of the primordial fluctuations
power spectrum (PS)\footnote{ The PS of density fluctuations is
  $P(\vec{k})=\left<|\delta_\rho(\vec{k})|^2\right>$, where
  $\delta_\rho(\vec{k})$ is the Fourier Transform of the normalized
  fluctuation field $(\rho(\vec{r})-\rho_0)/\rho_0$ \cite{glass}.}
\cite{glass}.  To avoid confusion, note that in statistical physics
the term ``scale invariance'' is used to describe the class of
distributions which are invariant with respect to scale
transformations. For instance, a magnetic system at the critical point
of transition between the paramagnetic and ferromagnetic phase, shows
a two-point correlation function which decays as a non-integrable
power law, i.e. $\xi(r) \sim r^{-\gamma}$ with $0<\gamma<3$
(super-Poisson distribution in Eq.\ref{sigma5}). The meaning of
``scale-invariance'' in the cosmological context is therefore
completely different, referring to the property that the mass variance
at the horizon scale be constant (see below) \cite{glass}.

\subsection{Basic Properties} 

Matter distribution in cosmology is assumed to be a realization of
a {\it stationary} stochastic point process that is also spatially
uniform.  In the early universe the homogeneity scale $\lambda_0$
is of the order of the inter-particle distance, and thus negligible,
while it grows during the process of structure formation driven by
gravitational clustering.  The main property of primordial density
fields in the early universe is that they are super-homogeneous,
satisfying
Eq.\ref{sigma5} with $n=1$. This latter property was firstly
hypothesized in the seventies \cite{harrison,zeldovich} and it
subsequently gained in importance with the advent of inflationary
models in the eighties \cite{glass}. 

In order to discuss this property, let us recall that { the
  fluctuations in the early universe} are taken to have Gaussian
statistics and a certain PS.  Since fluctuations are Gaussian, the
knowledge of the PS gives a complete statistical description of the
fluctuation field.  In a FRW cosmology there is a fundamental
characteristic length scale, the horizon scale $R_H(t)$ that is simply
the distance light can travel from the Big Bang singularity $t=0$
until any given time $t$ in the evolution of the Universe. { This
  scale linearly grows with time}.  Harrison \cite{harrison} and
Zeldovich \cite{zeldovich} introduced the criterion that matter
fluctuations have to satisfy on large enough scales. This is named the
Harrison-Zeldovich criterion (H-Z); it can be written as \cite{book}
\be \sigma^2 (R=R_H(t)) = {\rm constant}.
\label{H-Z-criterion}
\ee
This condition states that the mass variance at the horizon scale is
constant: it can be expressed more conveniently in terms of the PS for
which Eq.\ref{H-Z-criterion} is equivalent to assume $P(k) \sim k$
(the H-Z PS) and that in a spatial hyper-surface $\sigma^2(R) \sim
R^{-4}$ \cite{glass,book}.

\subsection{Physical implications of super-homogeneity} 

In order to illustrate the physical implications of the H-Z condition,
one may consider { the gravitational potential fluctuations}
$\delta\phi(\vec{r})$, which are linked to the density fluctuations
$\delta\rho(\vec{r})$ via the gravitational Poisson equation:
$\nabla^2\delta\phi(\vec{r})=4\pi G \delta\rho(\vec{r})\,. $ From this
{ equation, transformed into Fourier space}, it follows that the PS
of the { gravitational potential fluctuations}
$P_{\phi}(k)=\left<|\delta\hat\phi(\vec{k})|^2\right>$ is related to
the density PS $P(k)$ through the equation $ P_{\phi}(k)\sim
\frac{P(k)}{k^4}$.  The H-Z condition, $P(k) \sim k$, corresponds
therefore to $P_{\phi}(k) \propto k^{-3}$, so that the variance of the
gravitational potential fluctuations, $\sigma_{\phi}^2(R) \approx
\frac{1}{2} P_{\phi}(k) k^3|_{k =R^{-1}} $, is constant with $k$
\cite{glass}.

The H-Z condition is a {\it consistency constraint} in the framework
of FRW cosmology. Indeed, the FRW is a cosmological solution for a
perfectly { spatially and statistically} homogeneous universe,
about which fluctuations represent inhomogeneous perturbations. If
density fluctuations obey to a different condition than
Eq.\ref{H-Z-criterion}, and thus $n<1$ in Eq.\ref{sigma5}, then the
FRW description {\em will always break down } in the past or future,
as the amplitude of the perturbations become arbitrarily large or
small.  Thus the super-homogeneous nature of primordial density field
is a fundamental property independently on the nature of dark
matter. This is a very strong condition to impose, and it excludes
even Poisson processes ($n=0$ in Eq.\ref{sigma5}) \cite{glass} for
which fluctuations in gravitational potential diverge at large scales.

\subsection{The two-point correlation function  and super-homogeneity} 

The super-homogeneity (or H-Z) condition corresponds to the { limit
  condition expressed by Eq.\ref{intcos}, which represents} another
way { to reformulate that} $\lim_{k \rightarrow 0} P(k) =
0$.  This means that there is a fine tuned balance between small-scale
positive correlations and large-scale negative
anti-correlations~\cite{glass,book}.

Various models of primordial density fields differ for the behavior of
the PS at large wave-lengths { which is determined by the specific}
properties hypothesized { for } the dark matter component. For
example, in the Cold Dark Matter (CDM) scenario, where elementary
non-baryonic dark matter particles have a small velocity dispersion,
the PS decays as a power law $P(k) \sim k^{-2}$ at large $k$. For Hot
Dark Matter (HDM) models, where the velocity dispersion is large, the
PS presents an exponential decay at large $k$. However at small $k$
they both exhibit the H-Z tail $P(k) \sim k$ which is indeed the
common feature of all density { fields} compatible with FRW
models. The scale $r_c \approx k_c^{-1}$ at which the PS shows the
turnover from the linear to the decaying behavior is fixed to be the
size of the horizon at the time of equality between matter and
radiation \cite{peacock}.

{ Correspondingly,} the correlation function $\xi(r)$ of CDM (HDM)
models { (see Fig.\ref{xitheo})} presents the following behavior:
it is positive at small scales (decaying as $\xi(r) \sim r^{-1}$ for
CDM and being almost flat for HDM), it crosses zero at $r_c$ and then
it is negative approaching zero as $-r^{-4}$ (in the region
corresponding to $P(k) \sim k$) \cite{book}.

\subsection{Baryonic acoustic oscillations} 

{ Let us now mention} the baryon acoustic oscillations (BAO) scale
\cite{ew}. The physical description which gives rise to these
oscillations is based on fluid mechanics and gravity: when the
temperature of the plasma was hotter than $\sim 10^3$ K, photons were
hot enough to ionize hydrogen so that baryons and photons can be
described as a single fluid.  Gravity attracts and compresses this
fluid into the potential wells associated with the local density
fluctuations. Photon pressure resists this compression and sets up
acoustic oscillations in the fluid. Regions that have reached maximal
compression by recombination become hotter and hence are now visible
as local positive anisotropies in the cosmic microwave background
radiation (CMBR), if the different $k-$modes are assumed to have the
same phase { (which is the central hypothesis in this context}).

For our discussion, the principal point to note is that while
$k-$oscillations are de-localized, { the real space correlation
  function $\xi(r)$ has a localized feature at the scale} $r_{bao}$
corresponding to the frequency of oscillations in $k$ space.  This
simply reflects that the Fourier Transform of a regularly oscillating
function is a localized function. Formally the scale $r_{bao}$
corresponds to a scale where a derivative of $\xi(r)$ is not
continuous \cite{book,bb}.

\subsection{Size of structures and characteristic scales}

\begin{figure}
\begin{center}
\includegraphics*[angle=0, width=0.8\textwidth]{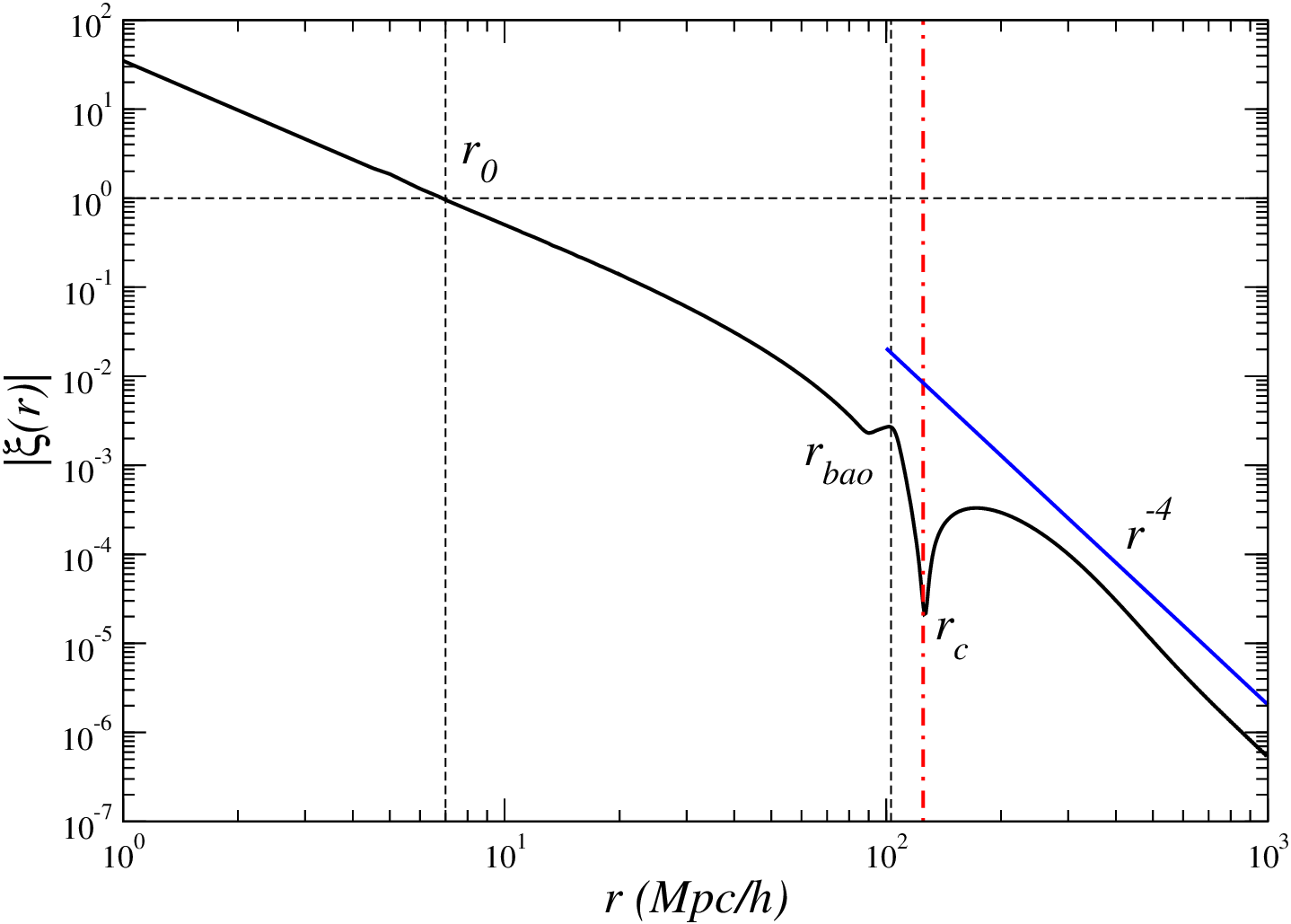}
\caption{Schematic behavior of the two-point correlation function for
  the LCDM case.  At small scales $r<r_0\approx 10$ Mpc/h (where
  $\xi(r_0)=1$) non-linear gravitational clustering has changed the
  initial shape of $\xi(r)$. At larger scales $\xi(r)$ has been only
  amplified by gravitational clustering in the linear regime. For
  $10<r<r_c\approx 120$ Mpc/h the correlation is positive and with
  small amplitude. At larger scales it is negative and characterized
  by the $\xi(r) \sim -r^{-4}$ behavior.  The location of $r_{bao}$ is
  fixed by cosmological parameters: in the example shown $r_{bao} <
  r_c$ as predicted by the ``concordance model'' \cite{eisenstein}. }
\label{xitheo}
\end{center}
\end{figure}

{ In summary, there are} three characteristic scales in  LCDM-type models
(see Fig.\ref{xitheo}).
The first is the homogeneity scale which depends on time $\lambda_0
= \lambda_0(t)$, the second is the scale $r_c$ where $\xi(r_c)=0$
(that is roughly proportional to the scale signing an exponential
decay of $\xi(r)$) which is fixed by the initial properties of the
matter density field, { which also determines } the third scale
$r_{bao}$.  { When} the homogeneity scale is smaller than
$r_{bao},r_c$, these two scales are substantially unchanged by
gravitational dynamics { as this is in the linear regime}. The rate
of growth of the homogeneity scale can be simply computed by using the
linear perturbation analysis of a self-gravitating fluid in an
expanding universe \cite{pee80}.  Given the initial amplitude of
fluctuations and the assumed initial PS of matter density
fluctuations, { under typical assumptions one finds that }
$\lambda_0(t_{now}) \approx 10$ Mpc/h \cite{springel}.

{ By characterizing } the two-point correlation function of
galaxy distribution we can identify three fundamental tests of
standard models
\footnote{For the power-spectrum there are additional complications,
  related how galaxies are biased with respect to the underlying
  density field: see \cite{bias,cdm_theo,ruth2011} for further
  details.}: 
\begin{itemize}
\item If the homogeneity scale $\lambda_0$ is much larger (i.e., a
  factor 5-10) than $\sim 10$ Mpc/h, then there is not enough time to
  form non-linear { large scale} structures in LCDM models \cite{sdss_aea}.
\item { If the   the zero crossing scale  of
  $\xi(r)$} is much larger than $\sim 100$ Mpc/h then there is a
  problem in the description of the early universe physics.
\item { A} clear test of inflationary models is given by the
  detection of the negative part of the correlation function, i.e. the
  range of scales it behaves as $\xi(r) \sim - r^{-4}$: all models
  necessarily predict such a behavior~\footnote{In the same range of
    scales the PS is expected to be linear with the wave-number,
    i.e. $P(k) \sim k$. However selection effects may change the
    behavior of the PS to constant but not the functional behavior of
    $\xi(r)$ \cite{bias,cdm_theo,ruth2011}.}.
\end{itemize}

\section{Testing assumptions in the statistical methods}
\label{stat} 

A number of different statistics, determined by making a volume
average in a finite sample, can be used to characterize a given
distribution. In addition, each statistical quantity can be measured
by using different estimators. For this reason we have to set up a
strategy to attack the problem if { a-priori} we do not know which
are the properties of the given finite sample distribution.  { In
  practice, to get the correct information from the data we have to
  reduce as much as possible the number of a-priori assumptions used
  the statistical methods.}

We limit our discussion to the case of interest, i.e. a set of $N$
point particles (i.e. galaxies) in a volume $V$. 
The microscopic number density can be simply written as 
$
n(\vec{r}) = \sum_i^N \delta^{3}(\vec{r}-\vec{r_i})  \,,  
$
where $\delta^{3}(\vec{r})$ is the Dirac delta function. The 
statistical quantities defined in Sect.\ref{prob} 
can be rewritten 
in terms of the stochastic variable
\be
\label{eq1} 
N_i(V) = \int_{V(\vec{y_i})} d^3x n(\vec{x}) \,,
\ee
where $\vec{y_i}$ identifies the coordinates of the center of the
volume $V$.  If the center $\vec{y_i}$ coincides with a point particle
position $\vec{r_i}$, then Eq.\ref{eq1} is a conditional
quantity. Instead, if the center $\vec{y_i}$ can be any point of space
(occupied or not by a particle) then the statistics in Eq.\ref{eq1} is
unconditional and it is useful to compute, for instance, the mass
variance defined in Eq.\ref{eq:sigma2}.

For inhomogeneous distributions, unconditional properties are
ill-defined (Sect.\ref{prob}) and thus we firstly analyze conditional
quantities to then pass, { only when} spatial homogeneity has been
detected inside the given sample, to { consider} unconditional
ones.  Therefore, in what follows { we take as volume $V$ in
  Eq.\ref{eq1} a sphere of radius $r$ centered in a distribution point
  particle}, i.e., we consider the stochastic variable defined by the
number of points in a sphere~\footnote{When we take a spherical shell
  instead of a sphere, then we define a differential quantity instead
  of an integral one.  } of radius $r$ centered on the $i^{th}$ point
of the given set, i.e.  $V=V(r;\vec{r_i})$.  The PDF $P(N(r)) =
P(N;r)$ of the variable $N_i(r)$ (at fixed $r$) contains, in
principle, information about moments of any order
\cite{saslawbook}. The first moment is the average conditional density
and the second moment is the conditional variance \cite{sdss_aea}.

However before considering the moments of the PDF we should study
whether they represent statistically meaningful estimates.  Indeed, in
the determination of statistical properties through volume averages,
one implicitly assumes that statistical quantities measured in
different regions of the sample are stable, i.e., that fluctuations in
different sub-regions are { actually} described by the same
PDF. Instead, it may occur that measurements in different sub-regions
show systematic (i.e., not statistical) differences, which depend, for
instance, on the spatial position of the specific sub-regions. In { 
  such a } case { the considered statistic is not stationary} in
space and its whole-sample average value (i.e., any finite-sample
estimation of the PDF moments) is not a meaningful descriptor.{ It
  is in this sense that it does not provide with a useful estimation
  of the ensemble average quantity}.

\subsection{Self-averaging} 
\label{sec:sstest}

A simple test to determine whether there are systematic finite size
effects affecting the statistical analysis in a given sample of linear
size $L$ consists in studying the PDF of $N_i(r)$ in sub-samples of
linear size $\ell < L$ placed in different spatial
regions of the sample { identified by their center-points} 
$\{S_1,...,S_N\}$.  When, at a given
scale $r<\ell$, $P(N(r),\ell;S_i)$ is the same, modulo statistical
fluctuations, in the different sub-samples, i.e., 
\be
\label{sstest}
P(N(r);\ell;S_i) \approx P(N(r);\ell;S_j) \; \forall i \ne j \;, 
\ee
it is possible to consider whole sample average quantities.  When
determinations of $P(N(r);\ell; S_i)$ in different regions $S_i$ show
{\it systematic} differences, then whole sample average quantities are
ill defined.  In general, this situation may occur because: (i) the
lack of the property of translational invariance or (ii) { the breaking
of self-averaging  property} due to finite-size
effects induced by large-scale structures/voids (i.e., long-range
correlated fluctuations).

While the breaking of translational invariance imply the lack of
self-averaging property the reverse is not true. For instance suppose
that the distribution is spherically symmetric, with origin at $r_*$
and characterized by a smooth density profile, function of the
distance from $r_*$ \cite{copernican}.  The average density in a
certain volume $V$, depends on the distance of it from $r_*$: there is
thus a systematic effect and Eq.\ref{sstest} is not satisfied. On the
other hand when a finite sample distribution is dominated by a single
or by a few structures then, even though it is translational invariant
in the infinite volume limit, a statistical quantity characterizing
its properties in a finite sample can be substantially affected by
finite size fluctuations.  For instance, a systematic effect is
present when the average (conditional) density largely differs when it
is measured into two disjointed volumes placed at different distances
from the relevant structures (i.e., fluctuations) in the sample. In
 a finite sample, if  structures are large enough, the measurements may
differ much more than a statistical scattering~\footnote{The determination
of statistical errors in a finite volume is also biased by finite size effects
\cite{cdm_theo,bao_aea}}. 
That systematic effect
sometimes is refereed to as cosmic variance \cite{saslaw2010} but that
is more appropriately defined as breaking of self-averaging properties
\cite{sdss_aea}, as the concept of variance (which involves already
the computation of an average quantity) maybe without statistical
meaning in the circumstances described above \cite{sdss_aea}.
In general, in the range of scales in which statistical quantities give
sample-dependent results, then they  do not represent 
fair estimations of asymptotic properties of the given
distribution \cite{sdss_aea}.

\subsection{Spatial homogeneity}

The self-averaging test (Eq.\ref{sstest}) is the first one to
understand whether a distribution is spatially homogeneous or not
inside a given sample. As long as the PDF $P(N,r)$ does not satisfy
Eq.\ref{sstest} then the distribution { is spatially inhomogeneous
  and} the moments of the PDF are not useful estimators of the
underlying statistical properties. Suppose that Eq.\ref{sstest} is
found to be satisfied up to given scale $r<L$.  Now we can ask the
question: { does the distribution become spatially homogeneous} for
$r<L$?

As mentioned in Sect.\ref{prob}, to { answer to this question} it
is necessary to employ statistical quantities that do not require the
assumption of spatial homogeneity, such as conditional ones
\cite{book,sdss_aea}.  Particularly the first moment of $P(N,r)$
provides an estimation of the average conditional density defined in
Eq.\ref{cond}, which can be simply written as
\be
\label{estimator_np} 
\overline{n(r)_p} =  \frac{1}{M(r)}
\sum_{i=1}^{M(r)} \frac{N_i(r)}{V(r)} = 
\frac{1}{M(r)}\sum_{i=1}^{M(r)} n_i(r) 
\;. 
\ee 
We recall that $N_i(r)$ gives the
number of points in a sphere of radius $r$ centered on the $i^{th}$
point and the sum is extended to the all $M(r)$ points contained in the
sample for which the sphere of radius $r$ is fully enclosed in the
sample volume (this quantity is $r$ dependent because of geometrical
constraints, see, e.g., \cite{sdss_aea}).  Analogously to
Eq.\ref{estimator_np} the estimator of the conditional variance can be
written as
\begin{equation}
  \label{estimator_condvar}
\overline{  \sigma_p^2(r)}  = \frac{1}{M(r)}
  \sum_{i=1}^{M(r)} n_i^2(r) - {\overline{n(r)_p}}^2 \;. 
\end{equation}
{ In the range of scales where} self-averaging properties are
satisfied, one may study the scaling properties of $\overline{n(r)_p}$
and of $\overline{ \sigma_p^2(r)}$.  As long as $\overline{n(r)_p}$
presents a scaling behavior as a function of spatial separation $r$,
as in Eq.\ref{scaling} with $D<3$, the distribution is spatially
inhomogeneous. When $\overline{n(r)_p} \approx$ const. then this
constant provides an estimation of the ensemble average density and
the scale $\lambda_0$, where the transition to a constant behavior
occurs, marks the homogeneity scale.  Only in this latter situation it
is possible to study the correlation properties of weak amplitude
fluctuations. This can be achieved by considering the function
$\xi(r)$ defined in Eq.\ref{2point}.

\subsection{The two-point correlation function}

Before proceeding, let us clarify some general properties of a generic
statistical estimator which are particularly relevant for the
two-point correlation function $\xi(r)$.  As mentioned above, in a
finite sample of volume $V$ we are only able to compute a statistical
estimator $\overline{X_V}$ of an ensemble average quantity $\langle X
\rangle$.  The estimator is valid if
\be 
\label{bias1}
\lim_{V\rightarrow
  \infty}  \overline{X_V} = \langle X \rangle \;.  
\ee 
If the ensemble average of
the finite volume estimator satisfies 
\be
\label{bias2} 
\langle \overline{X_V} \rangle = \langle X \rangle \ee the estimator
is unbiased. When Eq.\ref{bias2} is not satisfied then there is a
systematic offset which has to be carefully considered.  Note that the
violation of Eq.\ref{sstest} implies that Eq.\ref{bias2} is not valid
as well.  Finally the variance of an estimator is $\sigma_V^X =
\langle \overline{X_V}^2 \rangle - \langle \overline{X_V} \rangle^2 $.
The results given by an estimator must be discussed carefully
considering its bias and its variance in any finite sample. A strategy
to understand what is the effect of these features consists in
changing the sample volume $V$ and study finite size effects
\cite{book,cdm_theo,sdss_aea}.  This is crucially important for the
two-point correlation function $\xi(r)$ as any estimator $\overline{
  \xi(r)}$ { is generally biased}, i.e. it does not satisfy
Eq.\ref{bias2} \cite{cdm_theo,kerscher}.  { This occurs} because
the estimation of the { sample density} is biased when correlations
extend over the whole sample size, { or beyond it}. Indeed, the
most common estimator of the average density is
\be
\label{sd} 
\overline n = \frac{N}{V} \;, \ee where $N$ is the number of points in
a sample of volume $V$.  It is simple to show that its ensemble average
value can be written as \cite{cdm_theo} 
\be
\label{biasave}
\langle \overline n \rangle = \langle n \rangle \left( 1 +
\frac{1}{V} \int_V d^3 r \xi(r) \right) \;. 
\ee
Therefore only when $\xi(r) =0$ (i.e., for a Poisson
distribution), Eq.\ref{sd} is an unbiased estimator of the ensemble
average density: otherwise the bias is determined by 
the integral of the ensemble average correlation function over the volume $V$. 

The most simple estimator of $\xi(r)$  is the Full-Shell (FS) estimator 
\cite{cdm_theo} that can be simply written, by following the definition 
given in Eq.\ref{2point}, as 
\be
\label{xifs1}
\overline{ \xi(r)}  = \frac{\overline{ (n(r))_p}}{\overline n} -1 \,, 
\ee
where 
$\overline{ (n(r))_p}$ is the estimator of the conditional
density in spherical shells rather than
in spheres as for the case of Eq.\ref{estimator_np}. 
Suppose that in a spherical sample of radius $R_s$, to estimate the
sample density, instead of Eq.\ref{sd}, we use the estimator 
\be
\label{ne1}
\overline n= \frac{3}{4\pi R_s^3} \int_0^{R_s} \overline{ (n(r))_p}
4\pi r^2 dr \;.  
\ee 
Then, { by construction} the estimator defined
in Eq.\ref{xifs1} must satisfies the following integral constraint
\be
\label{xifs}
\int_0^{R_s} \overline{ \xi(r)} r^2 dr = 0 \;.  
\ee 
This condition is {\it satisfied independently of the functional shape
  of the underlying correlation function $\xi(r)$.}  Thus the integral
constraint for the FS estimator does not simply introduce an offset,
but it causes a change in the shape of $\overline{\xi(r)}$ 
for $r\rightarrow R_s$ . Other choices of the
sample density estimator \cite{cdm_theo,kerscher} and/or of the
correlation function introduce distortions similar to that in
Eq.\ref{xifs}.

In order to { show} the effect of the integral constraint for the
FS estimator, let us rewrite the ensemble average value of the FS
estimator (i.e., Eq.\ref{xifs1}) in terms of the ensemble average
two-point correlation function
\be
\label{estth}
 \langle \overline{ \xi(r)}\rangle 
= \frac{1+\xi(r)}{1+\frac{3}{R_s^3}\int_0^{R_s}
 \xi(r) r^2 dr} -1 \;. 
\ee
By writing Eq.\ref{estth} we assume that the stochastic noise is
negligible, { which, of course,} is not a good approximation at any
scale. However in this way we may be able to single out  the effect of
the integral constraint for the FS estimator. From Eq.\ref{estth} it
is clear that this estimator is biased, as it does not satisfy
Eq.\ref{bias2} but only Eq.\ref{bias1}. 

\begin{figure}
\begin{center}
\includegraphics*[angle=0, width=0.8\textwidth]{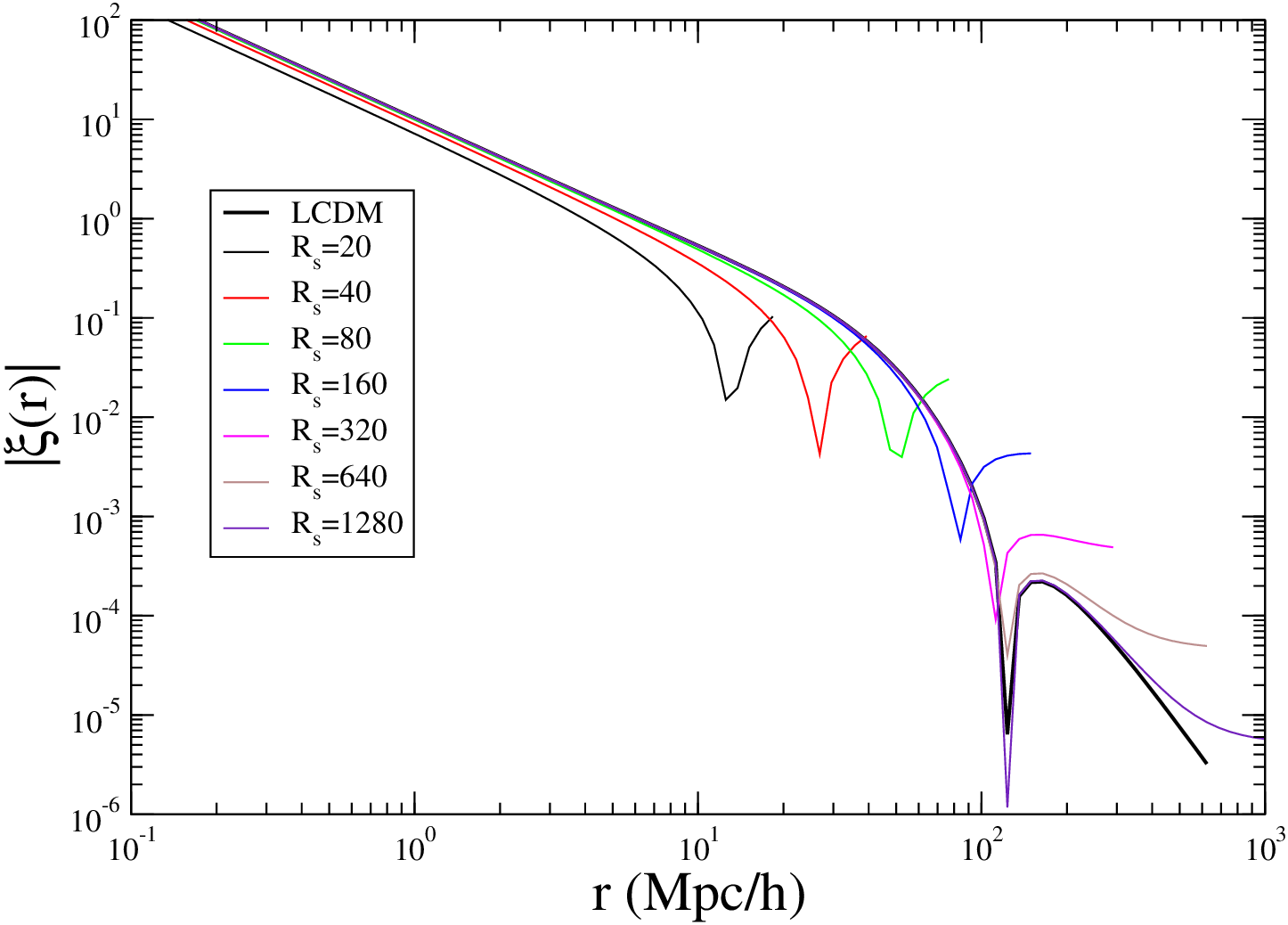} 
\end{center}
\caption{Absolute value of the estimation of the correlation function
  of the LCDM model with the integral constraint described by
  Eq.\ref{estth}. The tick solid line represents the theoretical
  model. The zero crossing scale correspond to the cusp ({ Adapted
    from \cite{cdm_theo})}.}
\label{figicvanilla}
\end{figure}

As an illustrative example, let us now consider the case in which the
theoretical $\xi(r)$ is a { given by the LCDM model}. The (ensemble
average) estimator given by Eq.\ref{estth}, in spherical samples of
different radius $R_s$, is shown in Fig.\ref{figicvanilla}.  One may
notice that for $R_s>r_c$ the zero point of $\overline{\xi(r)}$
remains stable, while when $R_s <r_c$ it { linearly grows with}
$R_s$.  The negative tail continues to be non-linearly distorted even
when $R_s>r_c$. For instance, when $R_s\approx 600$ Mpc/h we are not
able to detect the $\xi(r) \sim - r^{-4}$ tail that becomes marginally
visible only when $R_s > 1000$ Mpc/h.
Thus the stability of the zero-point crossing scale should be the
first problem to be considered in the analysis of $ \overline{ \xi(r)}
$, { clearly, once spatial homogeneity has been already proved}.


\section{Results in the data} 
\label{results} 

We briefly review the main results obtained by analyzing several
samples of the Sloan Digital Sky Survey (SDSS)
\cite{sdss_2007,sdss_epl,sdss_aea,gumbel,copernican} and of the Two
degree Field Galaxy Redshift Survey (2dFGRS)
\cite{2df_2006,2df_epl,2df_aea}.  In both catalogues we selected, in the
angular coordinates, a sky region such that (i) it does not overlap
with the irregular edges of the survey mask and (ii) it covers a
contiguous sky area.  We computed the metric distance $ R(z; \Omega_m,
\Omega_\Lambda) $ from the redshift $z$ by using the cosmological
parameters $\Omega_m=0.25$ and $\Omega_\Lambda=0.75$.

The SDSS catalogue includes two different galaxy samples constructing
by using different selection criteria: the main-galaxy (MG) sample and
the Luminous Red Galaxy (LRG) sample. In particular, the MG sample is
a flux limited catalogue with apparent magnitude $m_r < 17.77$
\cite{strauss2002}, while the LRG sample was constructed to be
volume-limited (VL) \cite{lrgsample}.  A sample is flux limited when
it contains all galaxies brighter than a certain apparent flux
$f_{min}$. There is an obvious selection effect in that it contains
intrinsically faint objects only when these are located relatively
close to the observer, while it contains intrinsically bright galaxies
located in wide range of distances \cite{zehavietal02}. For this
reason one constructs a volume limited (VL) sample by imposing a cut
in absolute luminosity $L_{min}$ and by computing the corresponding
cut in distance $r_{max} \approx \sqrt{L_{min}/(4\pi f_{min})}$, so
that all galaxies with $L>L_{min}$, located at distances $r<r_{max}$,
have flux $f>f_{min}$, and are thus included in the sample. By
choosing different cuts in absolute luminosity one obtains several VL
samples (with different $L_{min}, r_{max}$). Note that we use
magnitudes instead of luminosities and that the absolute magnitude
must be computed from the redshift by taking into account both the
assumptions on the cosmology (i.e. the cosmological parameters, which
very weakly perturb the final results given { the low redshifts
  involved, i.e., $z<0.2$}) and the K-corrections (which are measured
in the SDSS case).

For the MG sample we used standard K-corrections from the VAGC data
\cite{vagc}: we have tested that our main results do not depend
significantly on K-corrections and/or evolutionary corrections
\cite{sdss_aea}.  The MG sample angular region we consider is limited,
in the SDSS internal angular coordinates, by $-33.5^{\circ} \le \eta
\le 36.0^\circ$ and $-48.0^\circ \le \lambda \le 51.5^\circ$: the
resulting solid angle is $\Omega=1.85$ sr.  For the LRG sample, we
exclude redshifts $z>0.36$ and $z<0.16$ (where the catalogue is known 
be incomplete \cite{strauss2002,kazin}), so that the distance limits
are: $R_{min} = 465$ Mpc/h and $R_{max} = 1002$ Mpc/h. The limits in
R.A $\alpha$ and Dec. $\delta$ considered are: $\alpha \in [130^\circ,
  240^\circ]$ and $\delta \in [0^\circ,50^\circ]$. The absolute
magnitude is constrained in the range $M \in [-23.2,-21.2]$. With
these limits we find $N=41833$ galaxies covering a solid angle
$\Omega=1.471$ sr \cite{lrg_aea}.
Finally for 2dFGRS, to avoid the effect of the irregular edges of
the survey we selected two rectangular regions whose limits are
\cite{2df_aea}: in southern galactic cap (SGC)
($-33^\circ<\delta<-24^\circ$, $-32^\circ<\alpha<52^\circ$), and in
northern galactic cap (NGC) ($-4^\circ<\delta<2^\circ$,
$150^\circ<\alpha<210^\circ$); we determined absolute magnitudes $M$
using K-corrections from \cite{madgwick02,2df_aea}.


\subsection{Redshift selection function}

\begin{figure}
\begin{center}
\includegraphics*[angle=0, width=0.85\textwidth]{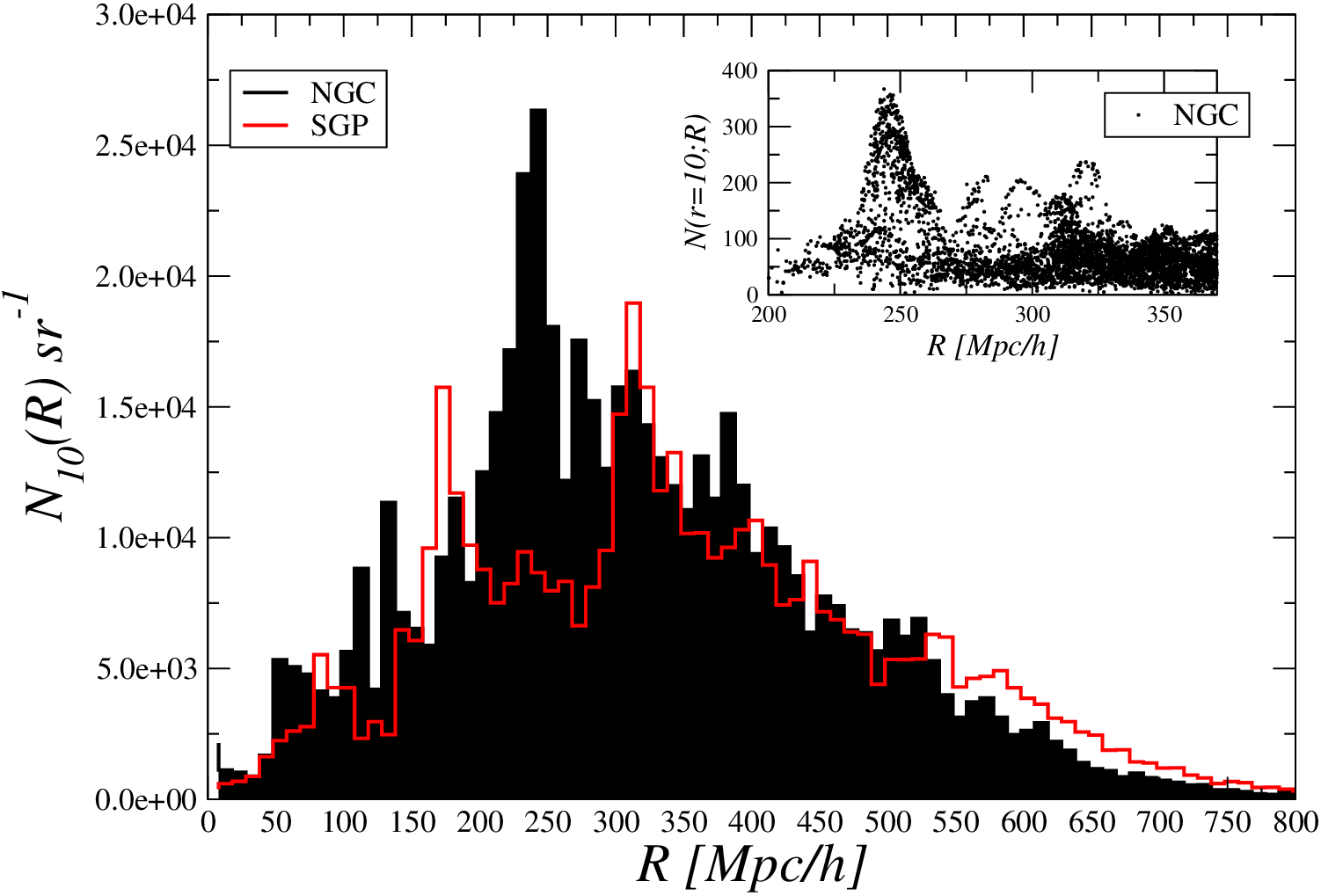}
\end{center}
\caption{Radial density { per unit solid angle} in bins of
  thickness 10 Mpc/h in the northern (NGC) and souther (SGC) part of
  the 2dFGRS magnitude limited sample.  There is a large structure at
  $\sim 240$ Mpc/h.  In the inset panel it is shown the distribution
  of $N_i(r;R)$ for $r=10$Mpc/h in a VL sample in the NGC.  (Adapted
  from \cite{2df_aea}).  }
\label{fig_counts_ml_ngp}
\end{figure}

In order to have a simple picture of the redshift distribution 
in a magnitude limited sample, we report Fig.\ref{fig_counts_ml_ngp}
galaxy counts as a function of the radial distance, in bins
of thickness 10 Mpc/h, in the northern and southern part of the 2dFGRS
\cite{2df_aea,2df_epl}.
One may notice that a sequence of structures and voids is clearly
visible, but there is an overall trend (a rise, a peak and then a
decrease of the density) which is determined by a luminosity selection
effect.  Indeed, $n(R)$ in a flux limited sample is usually called
redshift selection function, as it is determined by both the redshift
distribution and by the luminosity selection criteria of the survey.
It is thus not easy, by this kind of analysis, to determine, even at a
first approximation, the main properties of the galaxy distribution
in the samples. Nevertheless, one may readily compute that there is a
$\sim 30\%$ of difference in the sample density between the northern
and the southern part of the catalogue: one needs to refine the analysis
to clarify its significance.  Note that large scale $\sim 30\%$
fluctuations are not uncommon.  For instance, fluctuations have been
found in galaxy redshift and magnitude counts that are close to $50
\%$ occurring on $\sim 100$ Mpc/h scales
\cite{Ratcliffe98,busswell03,frith03,frith06}.


\subsection{Radial counts}

A more direct information about the value of the density in a VL
sample, is provided by the number counts of galaxies as a function of
radial distance $n(R)$ in a VL sample.  For a spatially homogeneous
distribution $n(R)$ should be constant while, for a fractal
distribution it should exhibit a power-law decay, even though large
fluctuations are expected to occur given that this not an average
quantity \cite{gsl01}.

\begin{figure}
\begin{center}
\includegraphics*[angle=0, width=1.\textwidth]{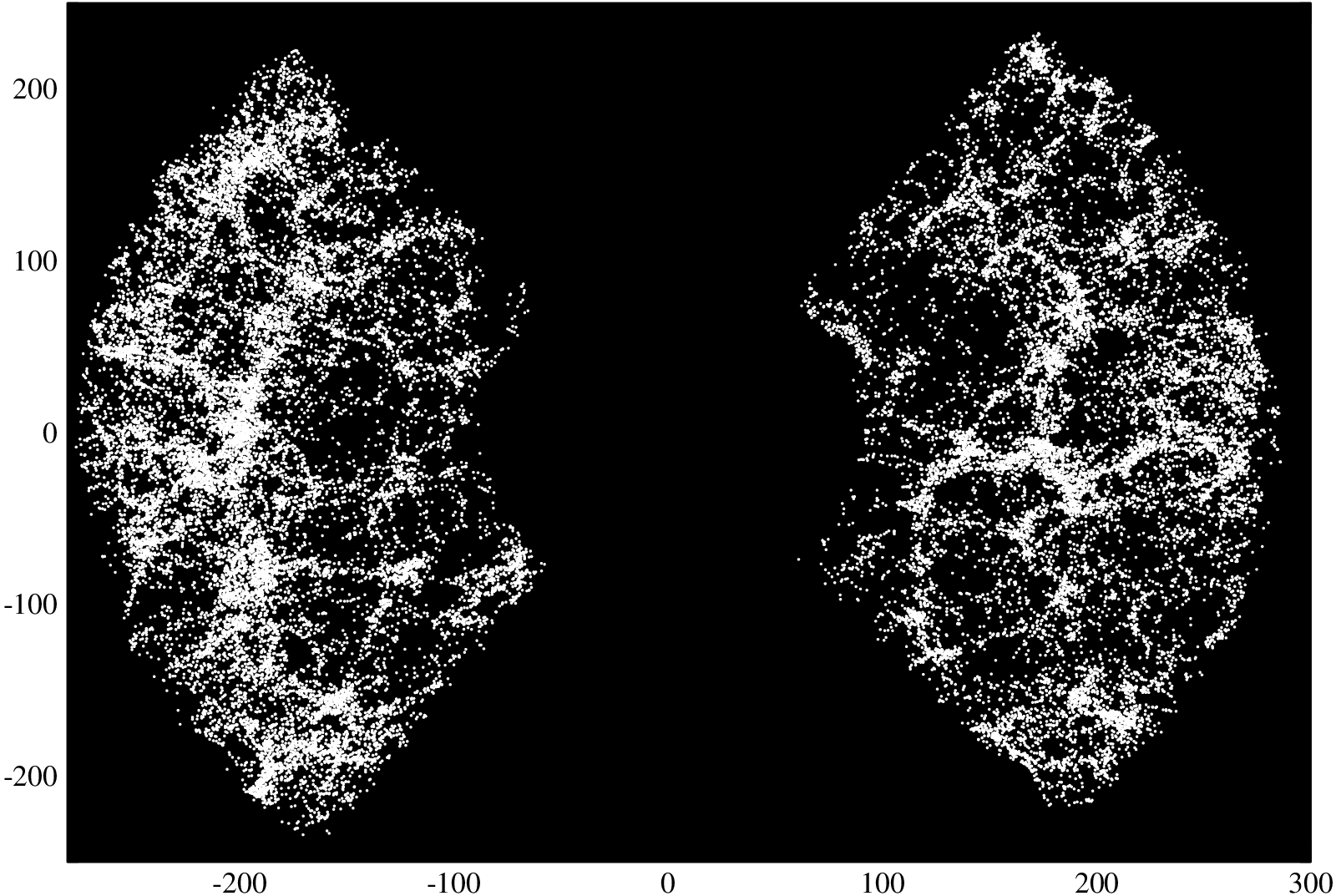}
\caption{Two different slices of the same SDSS MG VL sample.  Both
  slices cover an angular region of $100^\circ \times 10^\circ$, but
  in differnt directions.}
\label{fig3_a}
\end{center}
\end{figure}

\begin{figure}
\begin{center}
\includegraphics*[angle=0, width=0.45\textwidth]{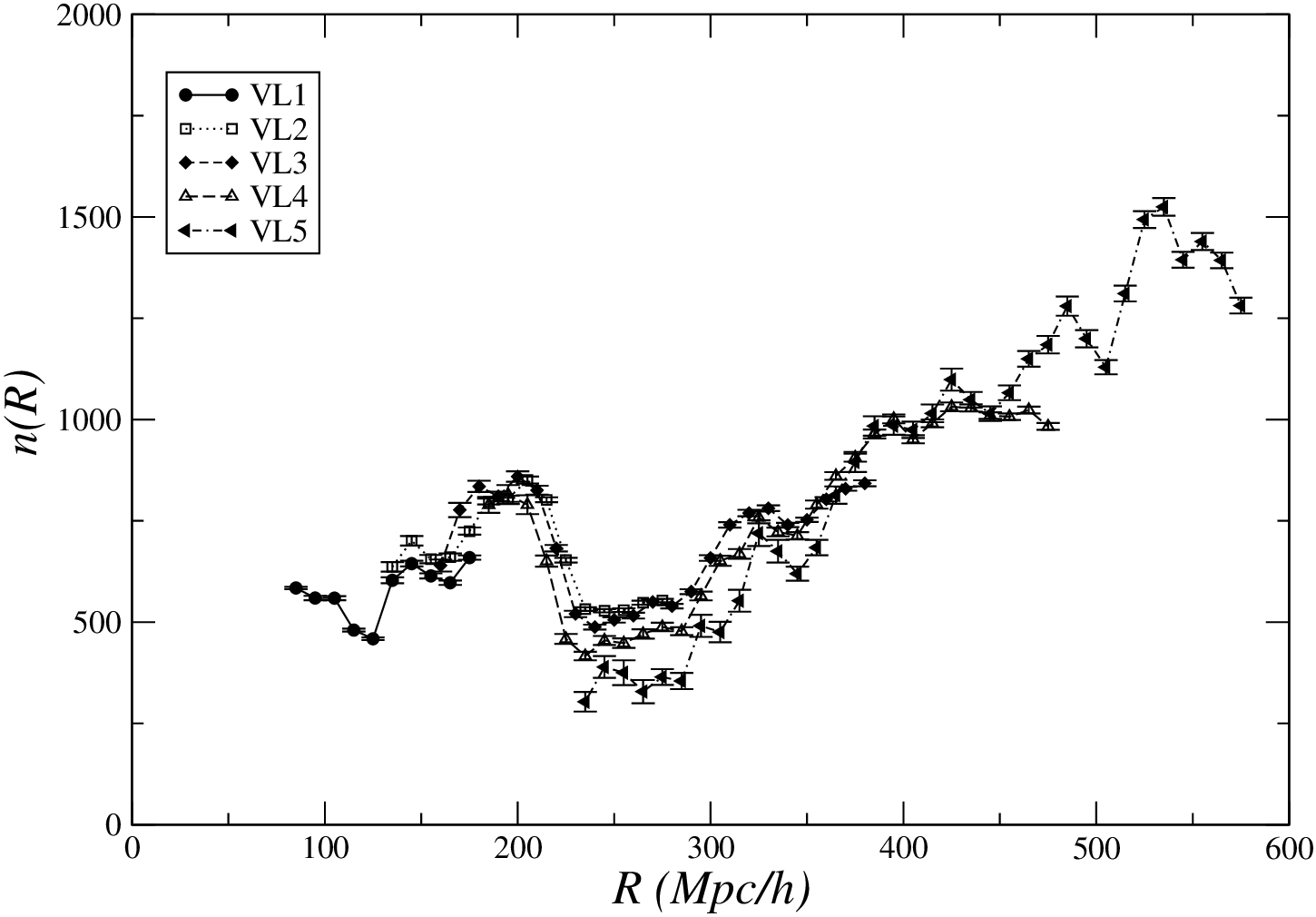}
\includegraphics*[angle=0, width=0.45\textwidth]{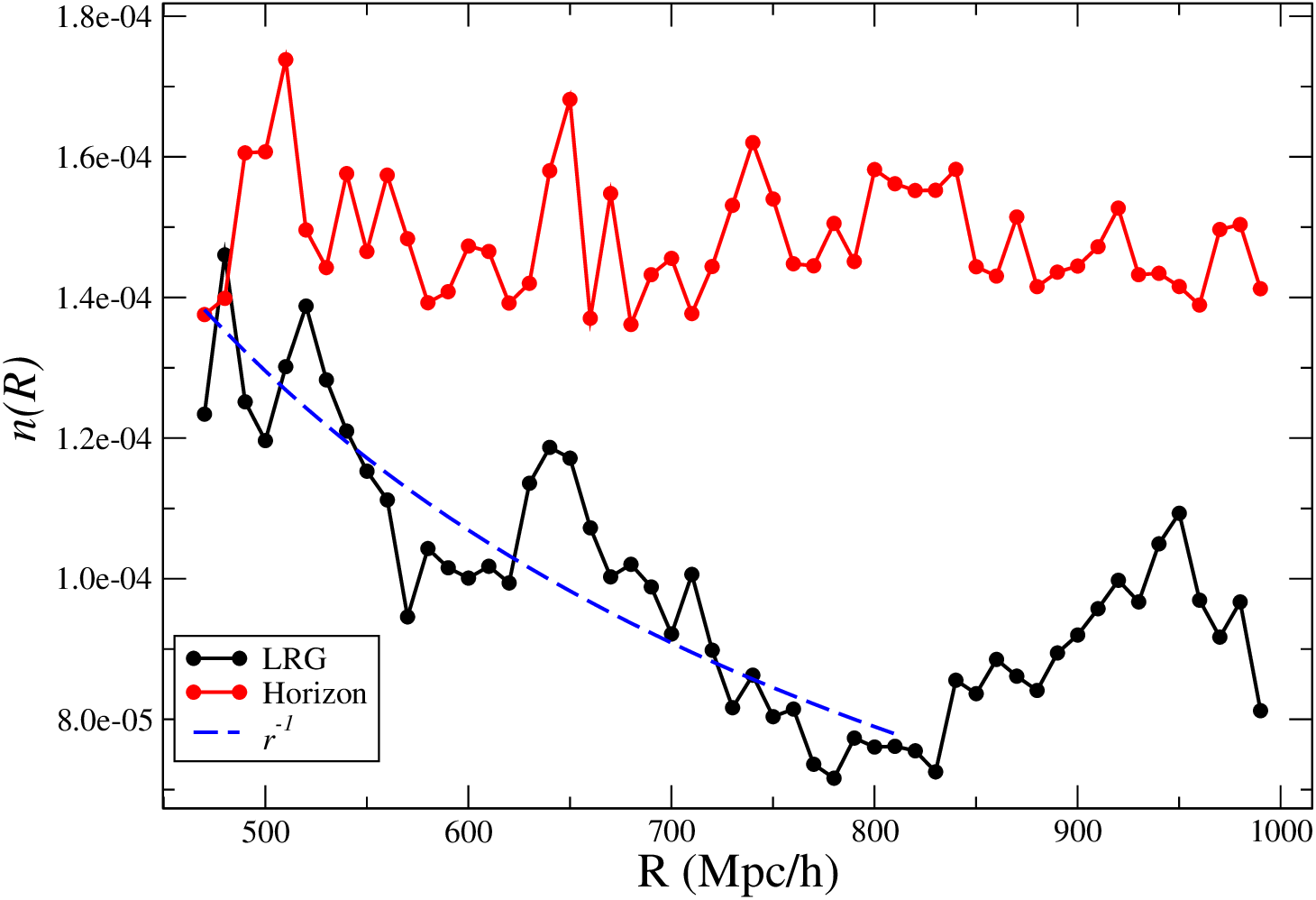}
\end{center}
\caption
{ {\it Left panel}: Radial density in the volume limited samples of
  the MG catalogue.  Note the amplitude of $n(R)$ for the MG VL samples
  has been normalized by taking into account the different selection
  in luminosity in the different samples ({ Adapted from
    \cite{sdss_aea})}.  {\it Right Panel}: The same for the LRG sample
  and for a mock sample extracted from the Horizon simulations
  \cite{horizon} (units are in (Mpc/h)$^{-3}$). The blue dashed line
  decays as $r^{-1}$ and it is plotted as reference.  ({ Adapted from
  \cite{lrg_aea})}.  }
\label{LRG+MGS_nr}
\end{figure}

In the SDSS MG VL samples, at small enough scales, $n(R)$ (see { 
  the left panel of} Fig.\ref{LRG+MGS_nr}) shows a fluctuating
behavior with peaks corresponding to the main structures in the galaxy
distribution \cite{sdss_aea}.  At larger scales $n(R)$ increases by a
factor 3 from $R\approx 300$ Mpc/h to $R \approx 600$ Mpc/h.  Thus
there is no range of scales where one may approximate $n(R)$ with a
constant behavior. The open question is whether the growth of $n(R)$
for $R>300$ Mpc/h is induced by structures { and/or by
  observational } selection effect in data: { in principle, both
  are possible.}  For instance in \cite{loveday} it is argued that a
substantial { galaxy }evolution causes that growth, while in
\cite{sdss_aea} { it is discussed that structures certainly
  contribute to the observed a behavior}. (Note that in mock
catalogues drawn from cosmological N-body simulations one measures an
almost constant density \cite{sdss_aea,2df_aea}).

Given that, by construction, also the LRG sample should be VL
\cite{lrg,zehavi,kazin} the behavior of $n(R)$ is expected to be
constant if galaxy distribution is close to uniform (up to Poisson
noise and radial clustering).  It is instead observed that the LRG
sample $n(R)$ shows an irregular and not constant behavior (see the
right panel of Fig.\ref{LRG+MGS_nr}) rather different from that { 
  found} in the MG sample.
Indeed, there are two main features: (i) a negative slope between
$400$ Mpc/h $<r<$ 800 Mpc/h (i.e., $0.16 < z < 0.28$) and (ii) a
positive slope up to a local peak at $r\sim 950$ Mpc/h (i.e., $z \sim
0.34$). Note that if $n(R)$ were constant we would expect a behavior
similar to the one shown by the mock sample extracted from the Horizon
simulation \cite{horizon} (see Fig.\ref{LRG+MGS_nr}) \cite{lrg_aea}.

An explanation that it is usually given { to interpret the behavior
  of $n(R)$} \cite{zehavi,kazin}, is that the LRG sample is ``quasi''
VL, { precisely because} it does not show a constant $n(R)$. Thus,
the { unexpected trends and features of} $n(R)$ are absorbed in the
properties of { the so-called ``the survey selection function''},
which is unknown {\it a priori}, but that is defined {\it a
  posteriori} as the difference between an almost constant $n(R)$ and
the behavior observed.  This explanation is unsatisfactory as it is
given {\it a posteriori} and no independent tests have been provided
to corroborate the hypothesis that an important observational
selection effect occurs in the data, other than the behavior of $n(R)$
itself. A different possibility is that the behavior of $n(R)$ is
determined, at least partially, by intrinsic fluctuations in the
distribution of galaxies and not by selection effects.

{ Note that, by} addressing the behavior of $n(R)$ to unknown
selection effects, it is implicitly assumed that more than the $20 \%$
of the total galaxies have not be measured for observational problems
\cite{lrg_aea}.  This looks improbable \cite{lrg} although a more
careful investigation of the problem must be addressed. Note also that
the deficit of galaxies would not be explained by a smooth
redshift-dependent effect, rather the selection must be strongly
redshift dependent as the behavior of $n(R)$ is not monotonic.  These
facts point, but do not proof, toward an origin of the $n(R)$ behavior
due to the intrinsic fluctuations in the galaxy distribution.

\subsection{Test on self-averaging properties} 

Galaxy counts provide only a rough analysis of fluctuations { as
  one is unable to compute a truly volume average quantity. In
  addition galaxy counts sample different scales differently as the
  volume in the different redshift bins is not the same}. The analysis
of the stochastic variable represented by the number of points in
spheres $N_i(r)$ an help to overcome these problems, as it is possible
to construct volume averages and because it is computed in a simple
real sphere sphere.  (See an example in the inset panel of
Fig.\ref{fig_counts_ml_ngp}).

\begin{figure}
\begin{center}
\includegraphics*[angle=0, width=0.85\textwidth]{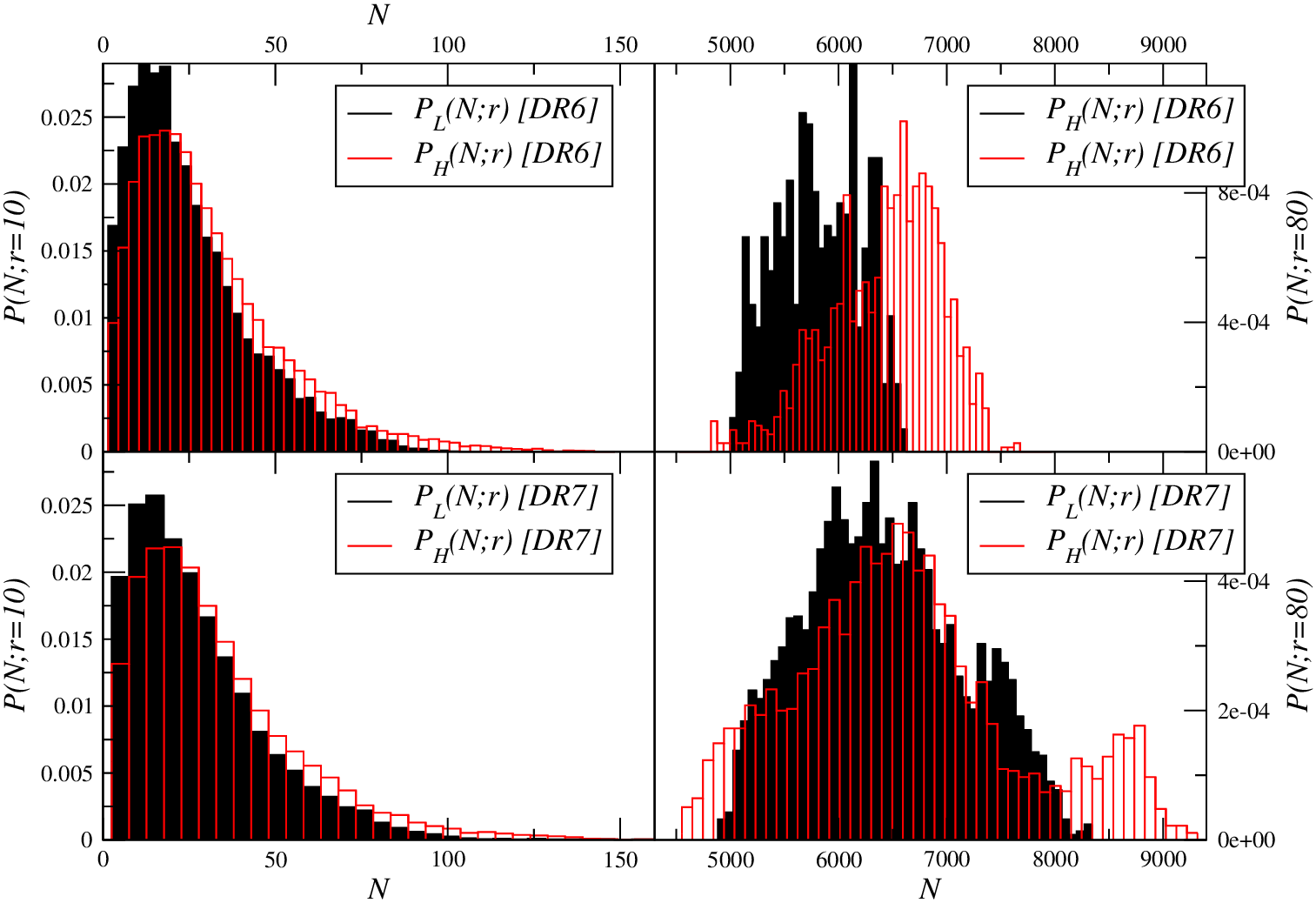}
\includegraphics*[angle=0, width=0.85\textwidth]{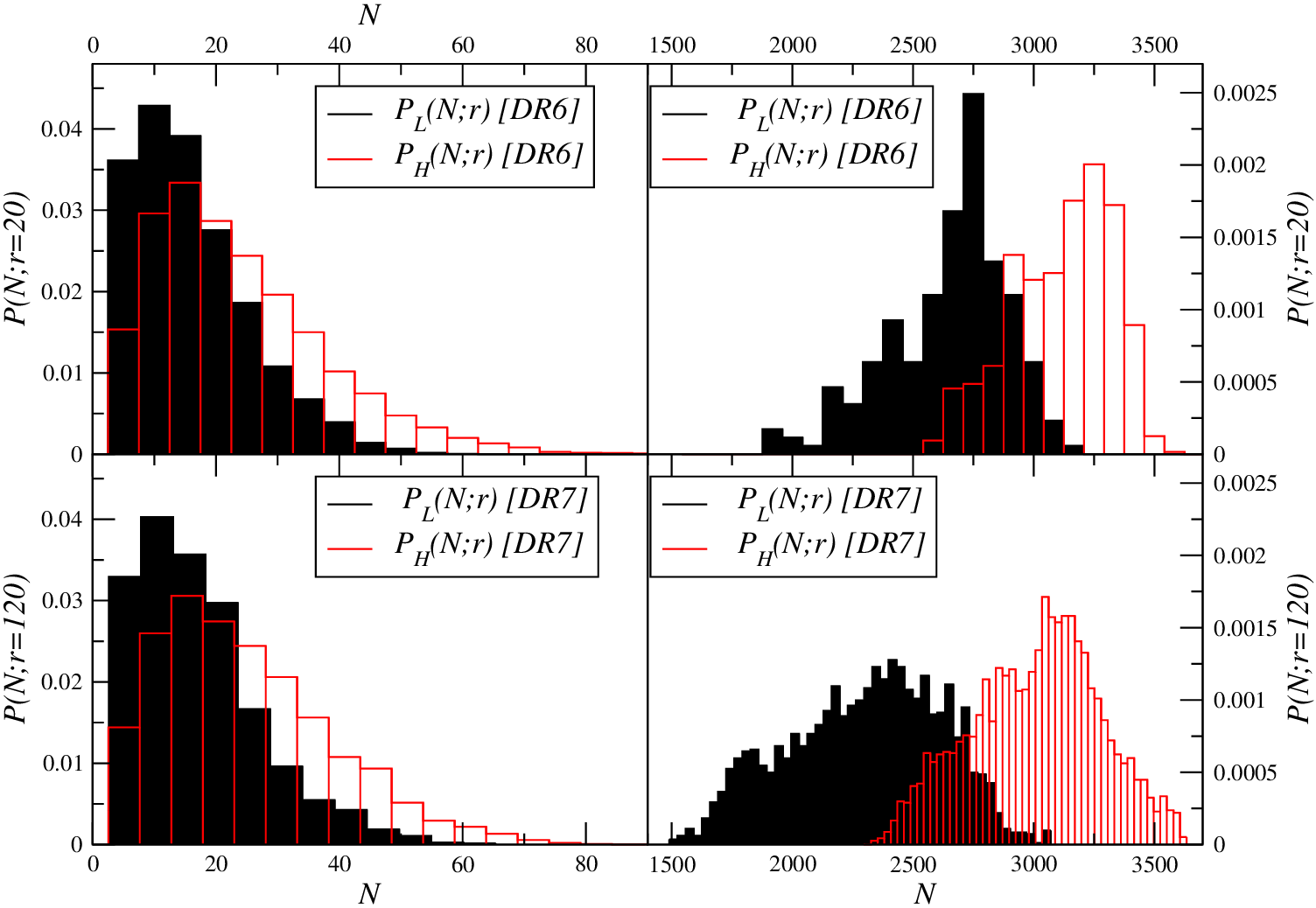} 
\end{center}
\caption{ {\it Upper Panels:} PDF of the counts in spheres in the
  sample defined by $R\in [125,400]$ Mpc/h and $M\in [-20.5,-22.2]$ in
  the DR6 and DR7 data, for two different values of the sphere radii
  $r=10$ Mpc/h and $r=80$ Mpc/h.  {\it Lower Panels:} The same but for
  the sample defined by $R\in [200,600]$ Mpc/h and $M\in
  [-21.6,-22.8]$ and for $r=20,120$ Mpc/h. (Adapted from
  \cite{copernican}).}
\label{fig:SS}
\end{figure}

Let us thus pass to the self-averaging test described in
Sect.\ref{sec:sstest}.  To this aim we divide the sample into two
non-overlapping regions of equal volume, one at low (L) and the other
at high (H) redshifts. We then measure the PDF $P_L(N;r)$ and
$P_H(N;r)$ in the two volumes ({ see \cite{copernican} for more
  details)}.  Given that the number of independent points is not very
large at large scales (i.e., $M(r)$ in Eq.\ref{estimator_np} not very
larger than $\sim 10^4$), in order to improve the statistics
especially for large sphere radii, we allow a partial overlapping
between the two sub-samples, so that galaxies in the L (H) sub-sample
count also galaxies in the H (L) sub-sample.  This overlapping clearly
can only smooth out differences between $P_L(N;r)$ and $P_H(N;r)$.

We first consider two SDSS MG VL samples from the data release 6 (DR6)
\cite{sdss_aea} and then from the DR7 \cite{copernican}.  In a first case
(upper - left panels of Fig.\ref{fig:SS}), at small scales ($r=10$
Mpc/h), the distribution is self-averaging (i.e., the PDF is
statistically the same) both in the DR6 sample (that covers a solid
angle $\Omega_{DR6} =0.94$ sr) than in the DR7 sample
($\Omega_{DR7}=1.85$ sr $\approx 2 \times \Omega_{DR6}$ sr).  Instead,
for larger sphere radii i.e., $r=80$ Mpc/h, (bottom - right panels of
Fig.\ref{fig:SS}) in the DR6 sample, the two PDF show clearly a
systematic difference. Not only the peaks do not coincide, but the
overall shape of the PDF is not smooth displaying a different
shape. Instead, for the sample extracted from DR7, the two
determinations of the PDF are in good agreement (within statistical
fluctuations).  We conclude that in DR6 for $r=80$ Mpc/h there are
large density fluctuations which are not self-averaging because of the
limited sample volume \cite{sdss_aea,copernican}.  They are instead
self-averaging in DR7 because the volume is increased by a factor two.

For the other sample we consider, which include mainly bright
galaxies, the breaking of self-averaging properties { occurs only}
for large $r$, { both in the DR6 and in the DR7 samples}.  { As
  mentioned above, radial distance-dependent selections, like galaxy
  evolution \cite{loveday}, could in principle give an effect in the
  same direction if they tend to increase the number density with redshift.
  However this would not change the main conclusion that, on large enough
  scales, self-averaging is broken}. Note that in the SDSS samples for
  small values of $r$ the PDF is found to be statistically stable in
  different sub-regions of a given sample.  For this reason we do not
  interpret the lack of self-averaging properties as due to a ``local
  hole'' around us: this would affect all samples and all scales,
  which is indeed not the case \cite{copernican}.  Because of these
  large fluctuations in the galaxy density field, self-averaging
  properties are well-defined only in a limited range of scales where
  it is then statistically meaningful to measure whole-sample average
  quantities \cite{sdss_aea,gumbel,copernican}.

\begin{figure}
\begin{center}
\includegraphics*[angle=0, width=0.35\textwidth]{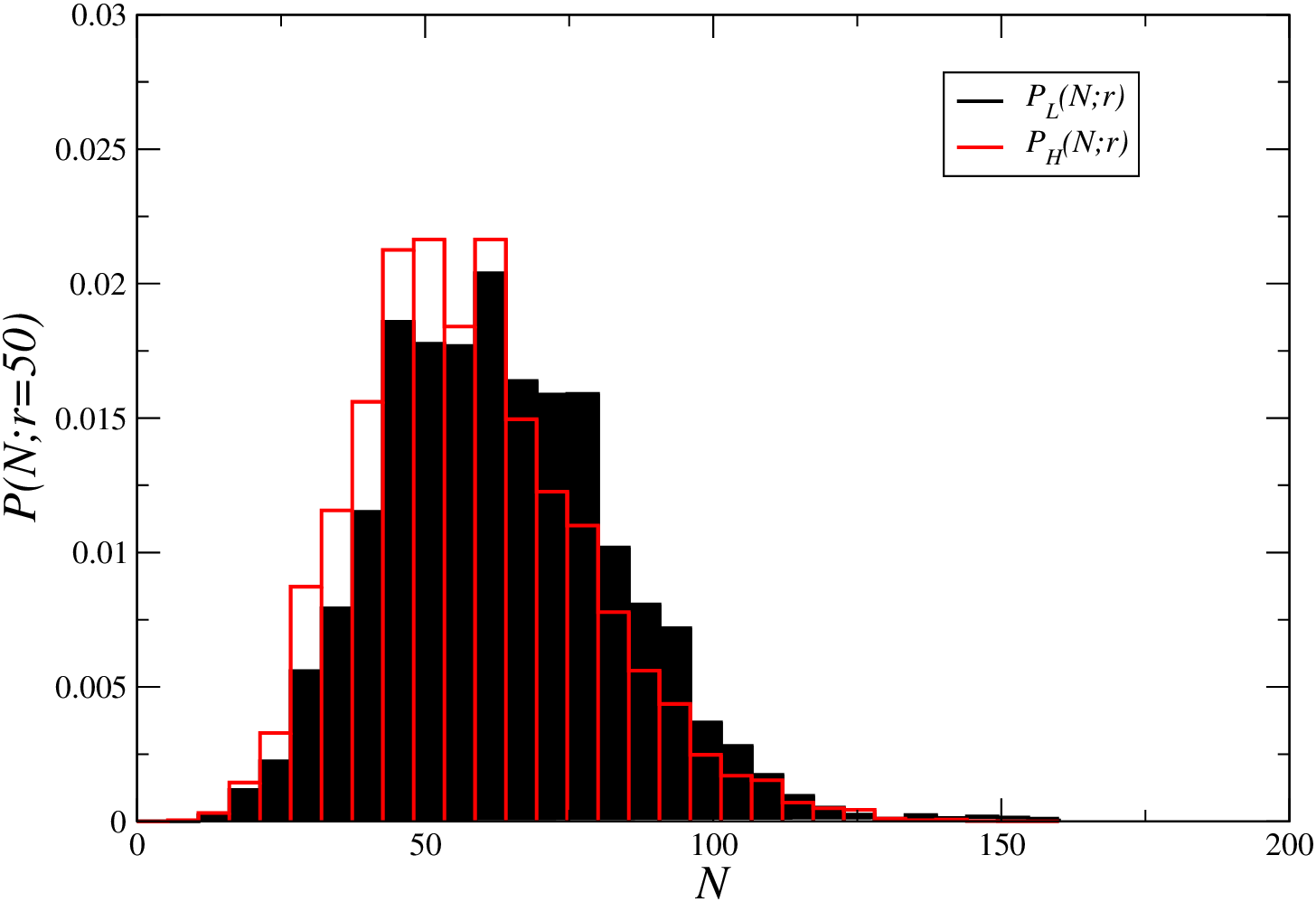}
\includegraphics*[angle=0, width=0.35\textwidth]{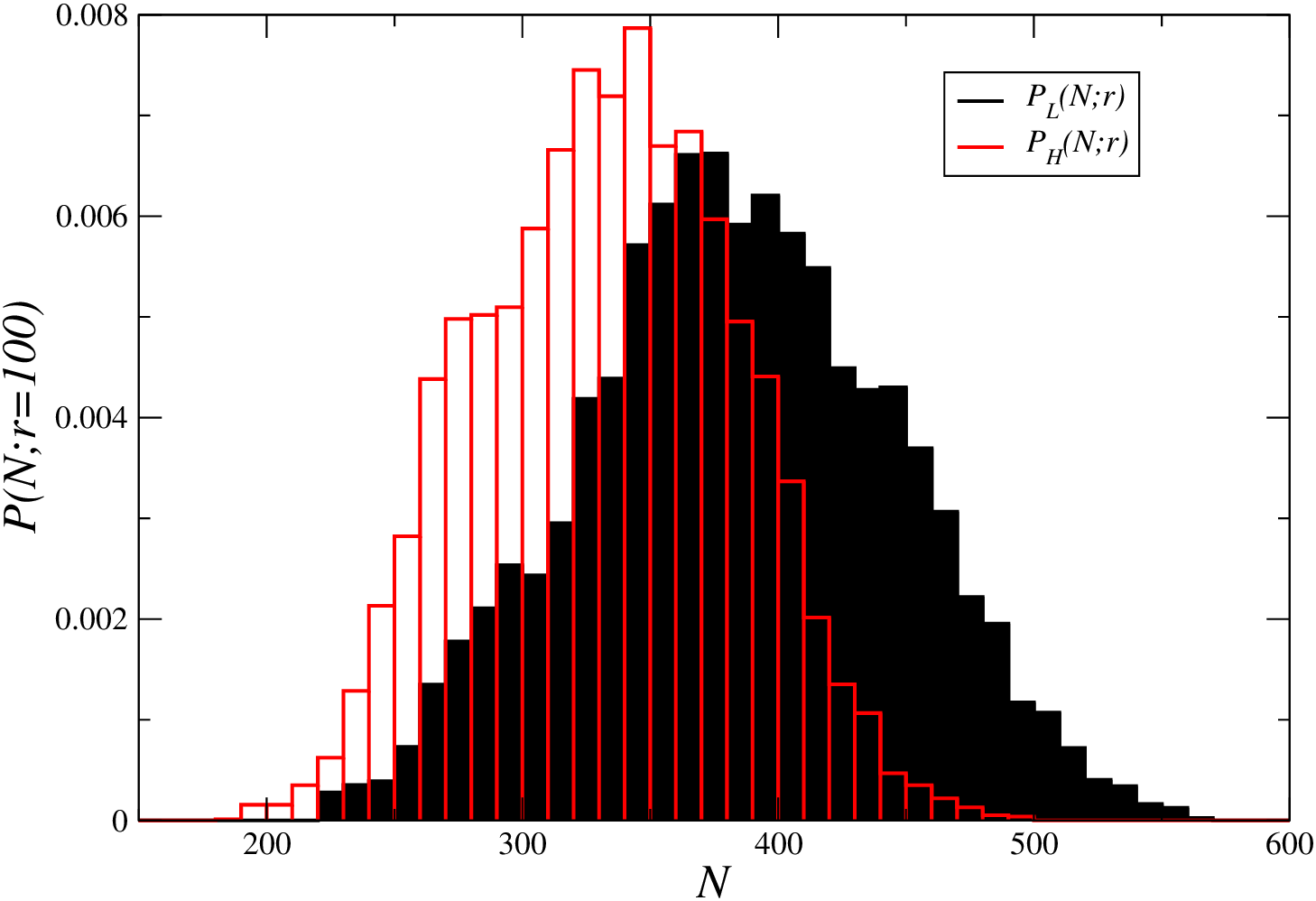}
\includegraphics*[angle=0, width=0.35\textwidth]{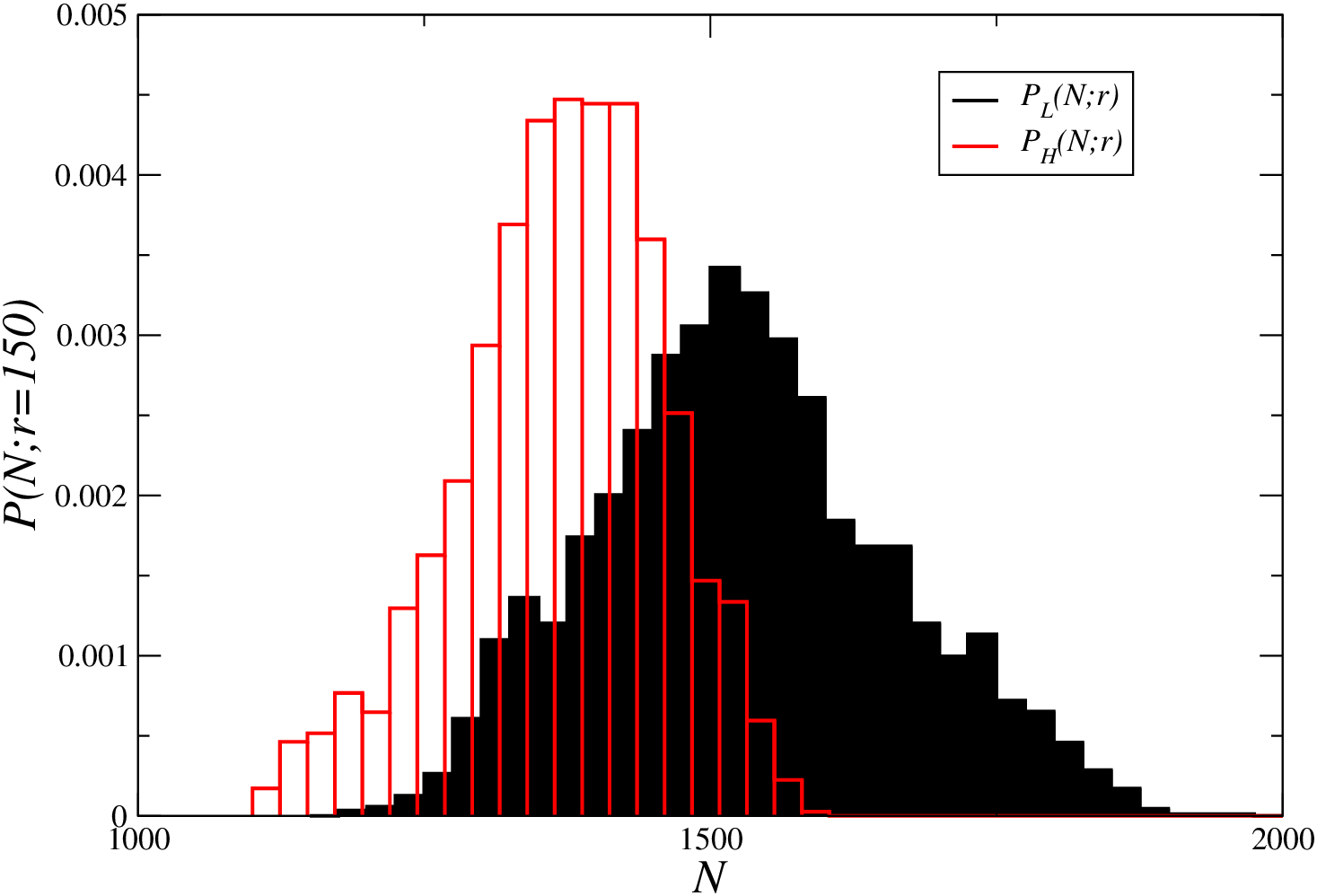}
\includegraphics*[angle=0, width=0.35\textwidth]{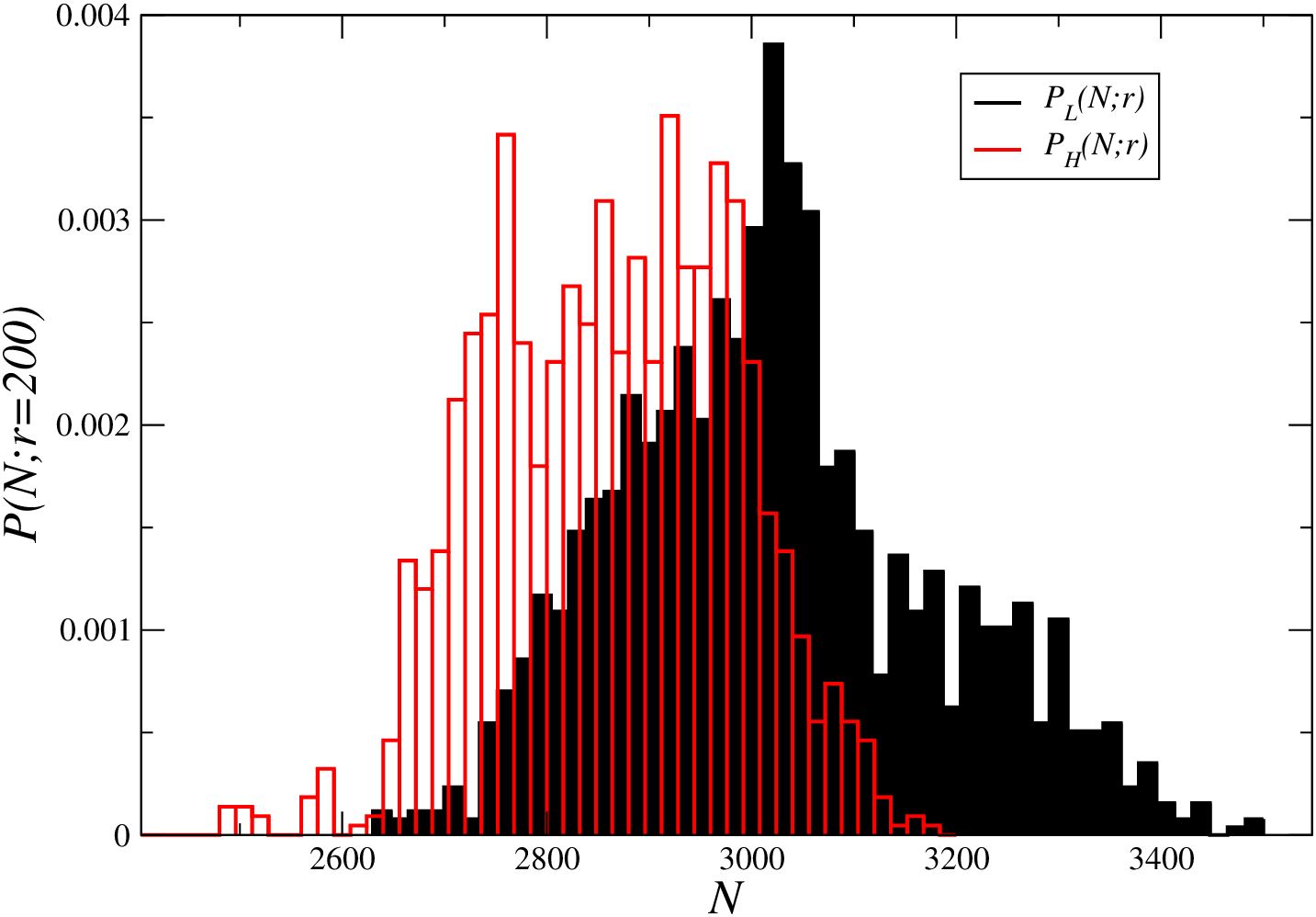}
\end{center}
\caption{ {\it Upper Left Panel:} PDF for $r=50$ Mpc/h in the LRG
  sample ({ Adapted from \cite{lrg_aea}). }}
\label{fig:SSLRG}
\end{figure}

For the LRG sample (see Fig.\ref{fig:SSLRG}) one may note that for
$r=50$ Mpc/h the determinations { in the two are separate parts of
  the sample} much closer than for lager sphere radii. { Indeed,
  fro $r>100$ Mpc/h there is actually a noticeable difference in the
  whole shape of the PDF}. The fact that $P_H(N;r)$ is shifted toward
smaller values than $P_L(N;r)$ is related to the decaying behavior of
the redshift counts (see Fig.\ref{LRG+MGS_nr}): most of the galaxies
at low redshifts see a relatively larger local density than the
galaxies at higher redshift.

{ In summary,} due the breaking of self-averaging properties in the
different samples for $r<150$ Mpc/h we conclude that there is no
evidence for a crossover to spatial uniformity.

\subsection{Probability density function and its moments} 

\begin{figure}
\begin{center}
\includegraphics*[angle=0, width=0.8\textwidth]{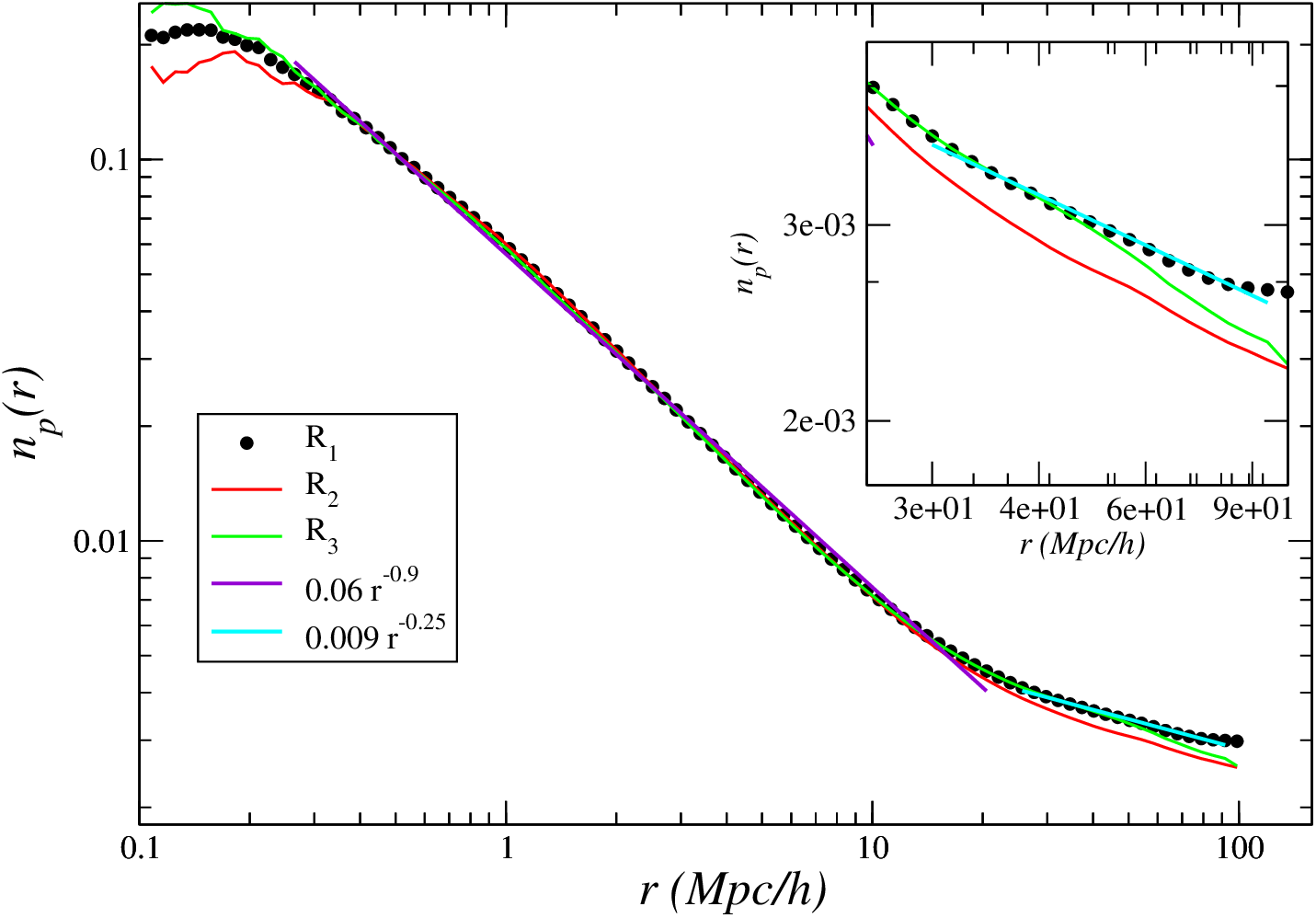}
\end{center}
\caption{  Conditional average density $\overline n(r)$ of galaxies
  as a function of radius ($R_1$). Note the change of slope at
  $\approx 20$ Mpc/h and note also that there is no flattening up to
  $\approx 100$ Mpc/h (in the inset panel it is shown a zoom at large
  scales).  The statistical significance of the last few points at the
  largest scales is weaker (see text).  The behavior of
  $\overline{n(r)} $ in two non-overlapping and equal volume regions,
  named $R_2$ and $R_3$, is also plotted.}
\label{fig:gammasigma}
\end{figure}

{ We can refine the analysis by characterizing the shape of the PDF
  and the scaling of its moments. In particular, in } the range of
scales { where} self-averaging properties are found to { be
  satisfied}, we can further characterize the shape of the PDF and the
scaling of its moments, { particularly the first moment, the
  behavior of the average conditional density (Eq.\ref{estimator_np})
  whose behavior is presented in Fig.\ref{fig:gammasigma}. In brief,
  it decays approximately as $r^{-1}$ up to $\approx 20$ Mpc/h where
  the decay changes to $\overline{n(r)} \approx 0.011 \times r^{-0.29}
  $} ~\footnote{ Alternatively, an almost indistinguishable fit is
  provided by a slow logarithmic one $\overline{n(r)} \approx
  \frac{0.0133}{\log r}$ \cite{gumbel}}. { Moreover, the density
  $\overline{n(r)}$ does not saturate to up to $\sim 100$ Mpc/h, i.e.,
  up to the largest scales probed in this sample where self-averaging
  properties have been tested to hold. In Fig.\ref{fig:gammasigma} it
  is also shown the behavior of $\overline{n(r)} $ into two
  non-overlapping regions of equal volume: these behaviors show the
  typical fluctuations affecting the estimation of this quantity. }
{ The scaling behavior of the conditional density implies that
  galaxy structures are characterized by non-trivial correlations for
  scales up to $r \approx 100$ Mpc/h, without a crossover towards
  spatial homogeneity.}

To probe the whole distribution of the conditional density $n_i(r)$,
we fitted the measured PDF with Gumbel distribution via its two
parameters $\alpha$ and $\beta$ \cite{gumbel}. The Gumbel distribution
is one of the three extreme value distribution
\cite{fisher28,gumbel58}. It describes the distribution of the largest
values of a random variable from a density function with faster than
algebraic (say exponential) decay. The Gumbel distribution's PDF is
given by
\begin{equation}
\label{gumbel}
 P(y)= \frac{1}{\beta} 
\exp\left[ - \frac{y-\alpha}{\beta} - 
\exp\left( - \frac{y-\alpha}{\beta} \right) \right] \;.
\end{equation}\
%
The mean and the variance of the Gumbel
distribution (Eq.\ref{gumbel}) is
$ \mu = \alpha + \gamma \beta, \quad 
 \sigma^2 = (\beta\pi)^2/ 6 $
where $\gamma=0.5772\dots$ is the Euler constant.

\begin{figure}
\begin{center}
\includegraphics*[angle=0, width=0.8\textwidth]{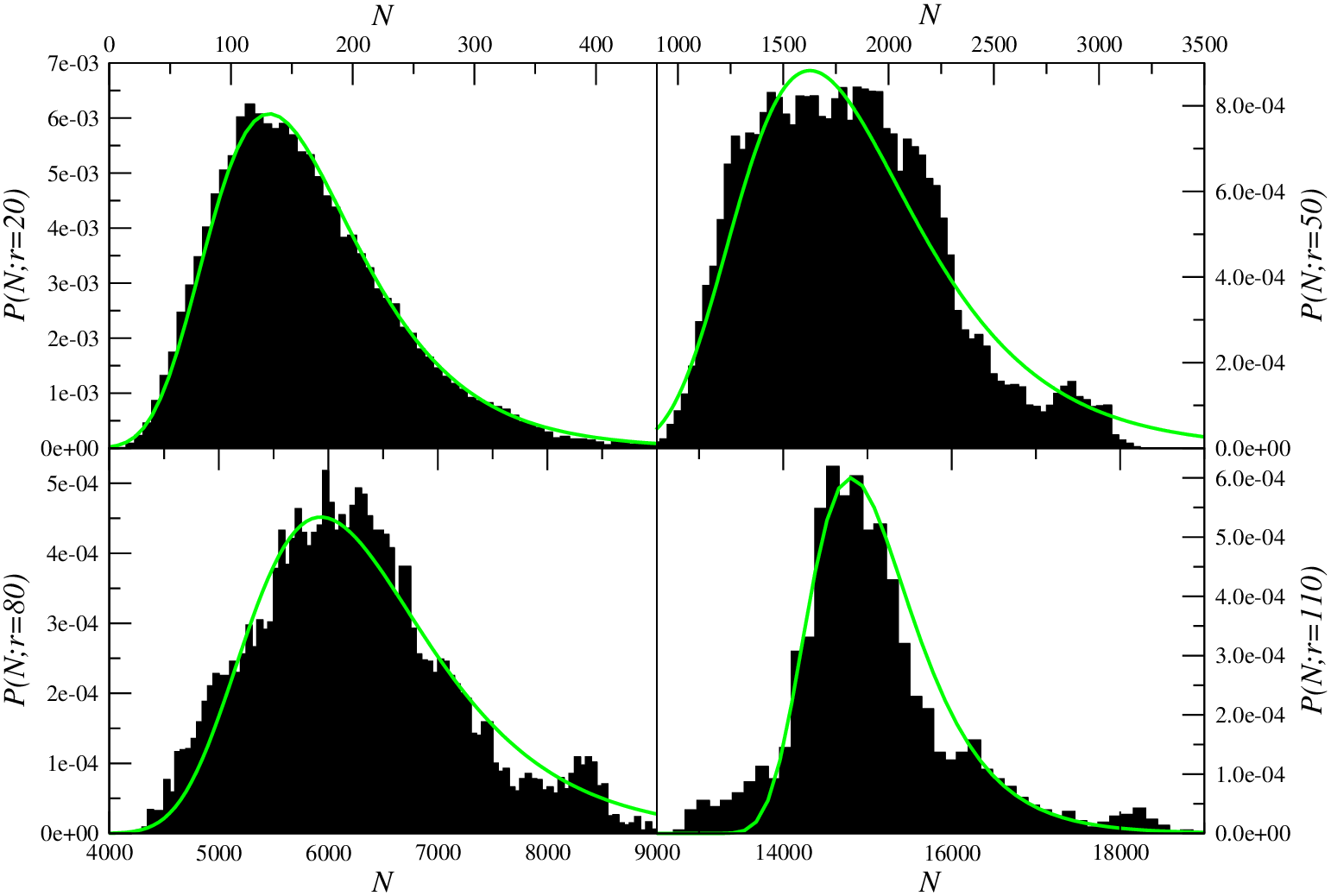}
\end{center}
\caption{ PDF for $r=20,50,80,100$ Mpc/h. The solid line
  corresponds to the best fit with a Gumbel distribution. }
\label{fig:gumbel}
\end{figure}

{ One of our best fit for the PDF is obtained for $r=20$ Mpc/h (see
  Fig.~\ref{fig:gumbel}).  At larger scales the fit get worst, but the
  Gumbel function remains a good fit even for $r=110$ Mpc/h. Given
  that the main source of uncertainty is, as discussed, finite volume
  systematic effects, it is not simple to determine the statistical
  significance of the Gumbel fit as systematic errors are larger than
  statistical ones. 

The fact that the PDF is clearly asymmetric, and well-fitted by a
Gumbel function, provides an additional evidence that correlations are
long-range. Indeed, due to the Central Limit Theorem, all homogeneous
point distributions with short-range correlations lead to Gaussian
fluctuations \cite{book}.  It was recently conjectured
\cite{bramwell09} that only three types of distributions appear to
describe fluctuations of global observables at criticality. In
particular, when the global observable depends weakly on the system
size (e.g., logarithmically), the corresponding distribution should be
a (generalized) Gumbel \cite{gumbel}.}


\subsection{Two-point correlation analysis} 

 When one determines the standard two-point correlation function one
 makes implicitly the assumptions that, inside a given sample the
 distribution is: (i) self-averaging and (ii) spatially uniform.  The
 first assumption is used when one computes whole sample average
 quantities. The second is employed when supposing that the estimation
 of the sample { density} gives a fairly good estimation of the
 ensemble average density.  When one of these assumptions, or both, is
 not verified then the interpretation of the results given by the
 determinations of the standard two-point correlation function must be
 reconsidered with great care.

\begin{figure}
\begin{center}
\includegraphics*[angle=0, width=0.4\textwidth]{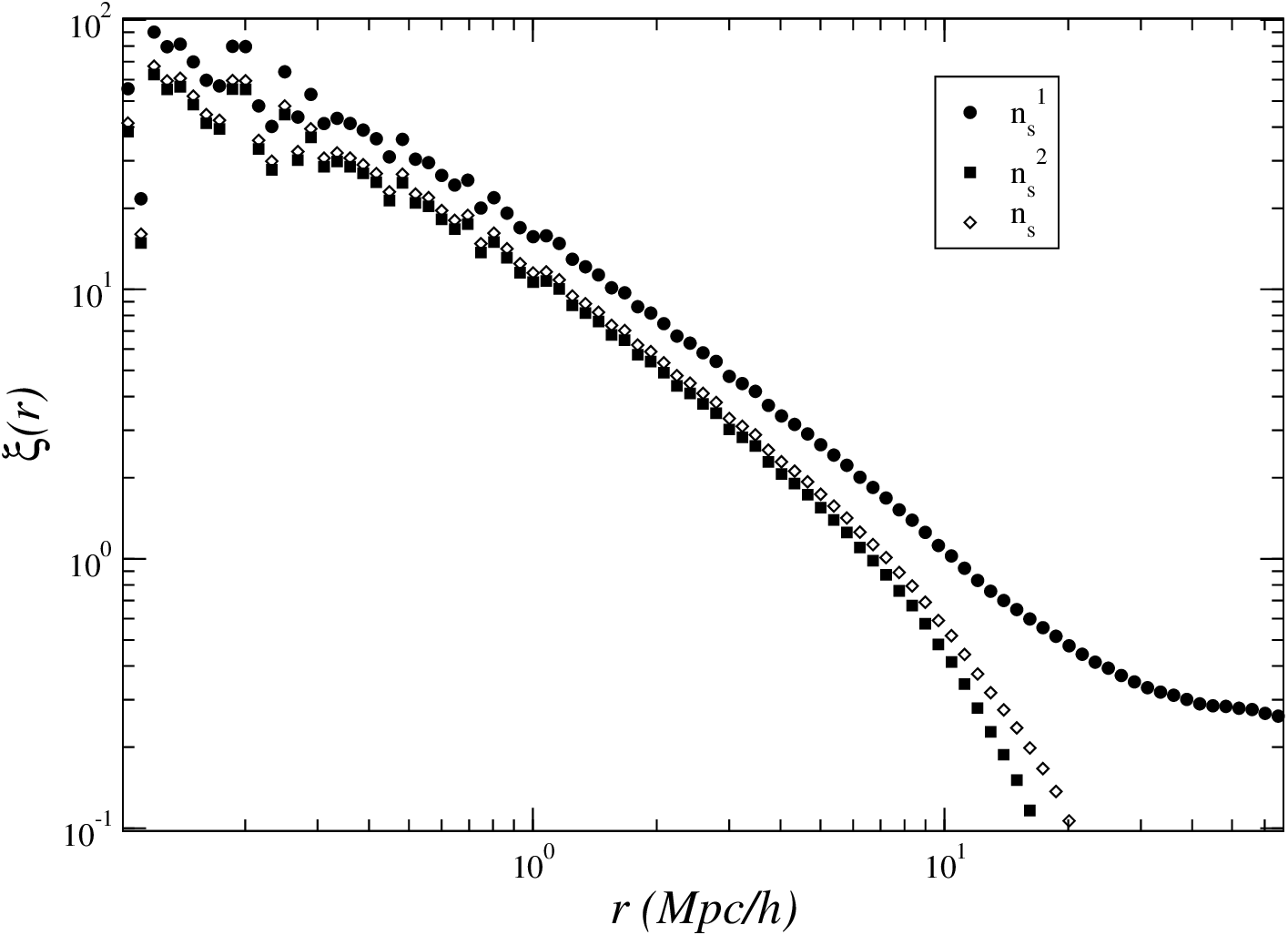}
\includegraphics*[angle=0, width=0.4\textwidth]{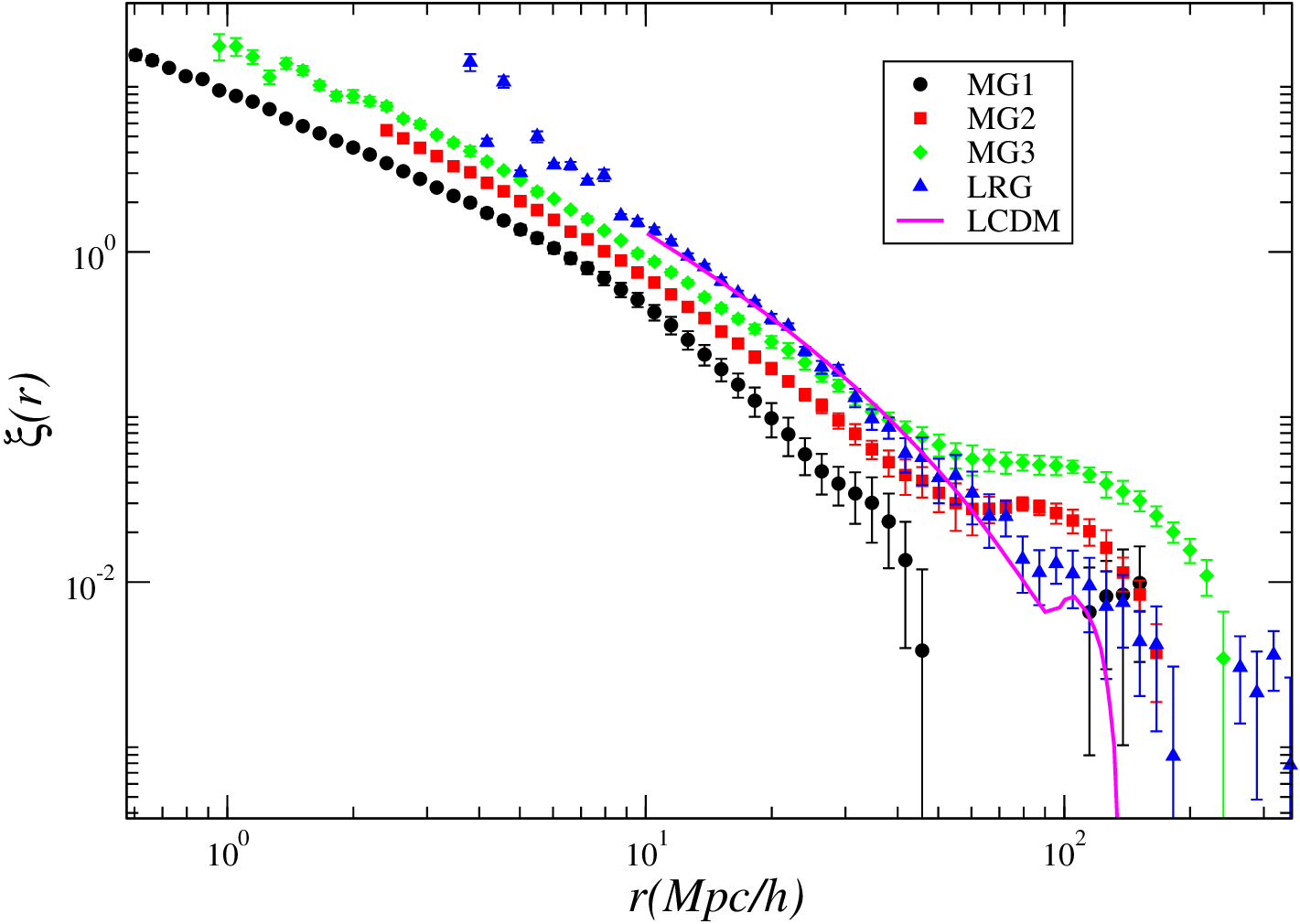} 
\end{center}
\caption{ {\it Left panel:} The two-point correlation function in a
  MG-VL sample estimated by Eq.\ref{xi2}: the sample average density
  is computed in spheres of radius $r^*=60$ Mpc/h and considering all
  center-points lying in a bin of thickness $\Delta R=50$ Mpc/h
  cantered at different radial distance $R$: $R_1=250$ Mpc/h ($n_s^1$)
  and $R_2=350$ Mpc/h ($n_s^2$).  The case in which we have used the
  estimation of the sample average $N/V$ ($n_s$) is also shown and it
  agrees with the FS estimator (adapted from \cite{sdss_aea}).  {\it
    Right Panel:} The Landy and Szalay \cite{ls} estimator of $\xi(r)$
  in various MG-VL sample and in a LRG sample of the SDSS. The most
  evident feature is the finite-size dependence of both the amplitude
  and the zero-crossing (adapted from \cite{bao_aea}). The solid line
  is a LCDM model.}
\label{figxiVL3}
\end{figure}

To show how non self-averaging fluctuations inside a given sample bias
the $\xi(r)$ analysis, we consider the estimator
\be 
\label{xi2}
\overline{\xi(r)} +1 = 
\overline{\xi(r;R,\Delta R)} +1 = \overline{n(r,\Delta r)_p}
 \cdot \frac{V(r^*)}{\overline{N(r^*;R,\Delta R)}} \,, 
\ee 
 where the second ratio on the r.h.s. is now the density of points in
spheres of radius $r^*$ averaged over the galaxies lying in a shell of
thickness $\Delta R$ around the radial distance $R$.  If the
distribution is homogeneous, i.e., $r^*>\lambda_0$, and statistically
stationary, Eq.\ref{xi2} should be (statistically) independent on the
range of radial distances $(R,\Delta R)$ chosen.  The two-point
correlation function is defined as a ratio between the average
conditional density and the sample average density: if both vary in
the same way when the radial distance is changed, then its amplitude
remains nearly constant. This however does not imply that the
amplitude of $\overline{\xi(r)}$ is meaningful, as it can happen that
the density estimated in sub-volumes of size $r^*$ show large
fluctuations and so the average conditional density, and this
occurring with a radial-distance dependence. { The
$\overline{\xi(r)}$ analysis gives a meaningful estimate of the
amplitude of fluctuations, {\it only if this amplitude
  remains stable by changing the relative position of the sub-volumes
  of size $r^*$ used to estimate the average conditional density and
  the sample average density}}. This is achieved by using the estimator
in Eq.\ref{xi2}. While standard estimators \cite{ls,kerscher,cdm_theo}
are not able to test for such an effect, as the main contributions for
both the conditional density and the sample average density come from
the same part of the sample (typically the far-away part where the
volume is larger).  We find large variations in the amplitude of
$\overline{\xi(r)}$ in the SDSS MG VL samples (see the left panel of
Fig.\ref{figxiVL3}).
This is simply an artifact generated by the large density fluctuations
on scales of the order of the sample sizes.  The results that the
estimator of $\xi(r)$ 
has nearly the same amplitude in different
samples, e.g.,
\cite{dp83,park,benoist,zehavi_earlydata,zehavi,norbergxi01,norbergxi02},
despite the large fluctuations of $N_i(r;R)$, are simply explained by
the fact that $\overline{\xi(r)}$ is a ratio between the average conditional
density and the sample average density: both vary in the same way when
the radial distance is changed and thus the amplitude is nearly
constant.

{ In the right panel of Fig.\ref{figxiVL3} it is plotted the
  behavior of $\overline{\xi(r)}$ in samples of different size.  This
  clearly show that there is a finite-size dependence of both the
  amplitude of the correlation function and of the zero-crossing
  scale.  Therefore the estimator of $\xi(r)$ is biased by
  volume-dependent systematic effects that make the detection of
  correlation amplitude only an estimate of their lower limit
  \cite{bao_aea}.}  A similar conclusion was reached by \cite{shanks},
i.e. that when corrections for possible systematics are taken into
account the correlation function may not be consistent with as high
amplitude a peak as claimed by \cite{eisenstein}.  To clarify this
issue, as discussed above, it is necessary to consider the set of
tests for statistical and spatial homogeneity discussed above.

Instead of investigating the origin of the fluctuating behavior of
$n(R)$, some authors \cite{kazin} focused their attention on the
effect of the radial counts on the determination of the two-point
correlation function.  In particular, they proposed mainly two
different tests to study what is the effect of $n(R)$ on the
determination of $\overline{\xi(r)}$.  The first test consists in
taking a mock LRG sample, constructed from a cosmological N-body
simulation of the LCDM model, and by applying a redshift selection
which randomly excludes points in such a way that the resulting
distribution has the same $n(R)$ of the real sample. Then one can
compare $\xi(r)$ obtained in the original mock and in redshift-sampled
mock. \cite{kazin} find that there is a good agreement between the
two.  This shows that the particular kind of redshift-dependent random
sampling considered for the given distribution, does not alter the
determination of the correlation function.  Alternatively we may
conclude that, under the assumption that the observed LRG sample is a
realization of a mock LCDM simulation, the $n(R)$ does not affect the
result. However, if we want to test whether the LRG sample has the
same statistical properties of the mock catalogue, we cannot clearly
proof (or disproof) this hypothesis by assuming a priori that this is
true.

In other words, standard analyses ask directly the question of whether
the data are compatible with a given model, by considering only a few
statistical measurements.  As it was shown by \cite{cg2011} the LRG
correlation function does not pass the null hypothesis, i.e. it are
compatible with zero signal, implying that the volume of current
galaxy samples is not large enough to claim that the BAO scale is
detected. In addition, by {\it assuming} that the galaxy correlations
are modelled by a LCDM model, one may find that the data allow to
constrain the position of the BAO scale. In our view this approach is
too narrow: in evaluating whether a model is consistent with the data,
one should show that {\it at least the main} statistical properties of
the model are indeed consistent with the data.  As discussed above, a
number of different properties can be considered, which are useful to
test the assumptions of (i) self-averaging { and} (ii) spatial
homogeneity.  When, inside the given sample, the assumption (i) and/or
(ii) are/is violated then the compatibility test of the data with a
LCDM model is not consistent with the properties of the data
themselves.


\section{Conclusion} 
\label{conclusion} 

The statistical characterization of galaxy structures presents a
number of subtle problems. These are associated both with the a-priori
assumptions which are encoded in the statistical methods used in the
measurements { of galaxy correlations } and in the a-posteriori
hypotheses that are invoked to explain certain measured behaviors.
{ These latter include for example, luminosity bias, galaxy
  evolution, observational selection effects, etc.}  { Therefore}
it is { necessary} to introduce direct tests to understand both
whether the a-priori assumptions are compatible with the data and
whether it is { justified} to introduce a-posteriori untested, but
plausible, hypotheses to interpret the results of the data analysis.
For instance, the analysis of the simple counts as a function of
distance, in the SDSS samples, shows clearly that the observed
behavior is incompatible with model predictions, { i.e., spatial
  homogeneity.}  As mentioned above, one may assume that the
differences between the model and the observations are due to
selection effects. Then this becomes clearly the most important
assumption in the data analysis that must be stressed clearly and
explicitly. In addition, one { must consider whether there is an
  independent way to study selection effects in the data.}

{ On the basis of the results have presented, aiming to directly
  test whether spatial and statistical homogeneity are verified inside
  the available samples we conclude that galaxy distribution is
  characterized by structures of large spatial extension. Given that
  we are unable to find a crossover towards homogeneity, the amplitude
  of these structures remain undetermined and their main
  characteristic is represented by the scaling behavior of their
  relevant statistical properties. In particular, we discussed that
  the average conditional density presents a scaling behavior of the
  type $ \sim r^{-\gamma}$ with $\gamma \approx -1$ up to $\sim 20$
  Mpc/h followed by a $\gamma \approx -0.3$ behavior up to $\sim 100$
  Mpc/h. Correspondingly the probability density function (PDF) of
  galaxy (conditional) counts in spheres shows a relatively long tail:
  it is well fitted by the Gumbel function instead than by the
  Gaussian function, as it is generally expected for spatially
  homogeneous, { short range correlated}, density fields.

{ The statistical tests introduced here can thus provide direct
  observational evidences, at small scales and low redshifts (when $z\ll
  1$ we can neglect the important complications of evolving
  observations onto a spatial surface for which we need a specific
  cosmological model) of the basic assumptions used in the derivation
  of the FRW models, i.e. spatial and statistical homogeneity. In this
  respect it is worthing to further clarify  the subtle
  difference between these two concepts \cite{copernican}.  The
    concordance model of the universe combines {\it three} fundamental
    assumptions: (i) Einstein's field equations to determine the
    dynamics of space-time. (ii) Statistical homogeneity and isotropy,
    i.e., that ``the Earth is not in a central, specially favored
    position'' \cite{bondi,pedro2}. This requirement can be though to
    be the Copernican Principle which is a fundamental principle
    because one wants to avoid any special point or direction. (iii)
    Spatial homogeneity: this requirement is not a fundamental one as
    (ii) but plays the crucial role of simplifying the solutions of the
    Einstein's field equations.

The Cosmological Principle is usually meant to include both the
requirement of statistical homogeneity and isotropy and of spatial
homogeneity: these assumptions are often simply summarized in the
requirement that the universe is homogeneous and isotropic.  However
one must bear in mind the fact that the universe looks the same, at
least in a statistical sense, in all directions and that all observers
are alike does not imply spatial homogeneity of matter distribution.}
It is however this latter condition that allows us to treat, above a
certain scale, the density field as a smooth function, a fundamental
hypothesis used in the derivation of the FRW metric.

We have shown that galaxy
distribution in different samples of the SDSS is compatible with the
assumption that this is transitionally invariant, i.e. it satisfies
the requirement of the Copernican Principle that there are no spacial
points or directions. 
}
On the other hand, we found that there are no
clear evidences of spatial homogeneity up to scales of the order of
the samples sizes, i.e. $\sim 100$ Mpc/h.  This implies that galaxy
distribution is not compatible with the stronger assumption of spatial
homogeneity, encoded in the Cosmological Principle. In addition, at
the largest scales probed by these samples (i.e., $r\approx 150$
Mpc/h) we found evidences for the breaking of self-averaging
properties, i.e. that the distribution is not statistically
homogeneous. Forthcoming redshift surveys will allow us to clarify
whether on such large scales galaxy distribution is still
inhomogeneous but statistically stationary, or whether the evidences
for the breaking of spatial translational invariance found in the SDSS
samples were due to selection effects in the data.

We note an interesting connection between spatial
inhomogeneities and large scale flows which can be hypothesized by
assuming that the gravitational fluctuations in the galaxy
distribution reflect those in the whole matter distribution, and that
peculiar velocities and accelerations are simply correlated.  Peculiar
velocities provide an important dynamical information as they are
related to the large scale matter distribution.  By studying their
local amplitudes and directions, these velocities allow us, in
principle, to probe deeper, or hidden part, of the Universe.  The
peculiar velocities are indeed directly sensitive to the total matter
content, through its gravitational effects, and not only to the
luminous matter distribution. However, their direct observation
through distance measurements remains a difficult task.  Recently,
there have been published a growing number of observations of
large-scale galaxy coherent motions which are at odds with standard
cosmological models
\cite{watkins2008,lavaux2008,kaslinski2008,kaslinski2009}.

It is possible to consider the PDF of gravitational force fluctuations
generated by source field represented by galaxies, and test whether it
converges to an asymptotic shape within sample volumes. In several
SDSS sample we find that density fluctuations at the largest scales
probed, i.e. $r\approx 100$ Mpc/h, still significantly contribute to
the amplitude of the gravitational force \cite{force_aea}. Under the
hypotheses mentioned above we may conclude that that large-scale
fluctuations in the galaxy density field can be the source of the
large scale flows recently observed.

{ From the theoretical point of view, it is then necessary to
  understand how to treat inhomogeneities in the framework of General
  Relativity
  ~\cite{pedro2,ellis,buchert,wiltshire,clarkson,sisky,marra,celerier}.}
To this aim one needs to carefully consider the information that can
be obtained from the data.  At the moment it is not possible to get
some statistical information for large redshifts ($z\approx 1$), but
the characterization of relatively small scales properties (i.e.,
$r<200$ Mpc/h) is getting more and more accurate.  According to FRW
models the linearity of Hubble law is a consequence of the homogeneity
of the matter distribution. Modern data show a good linear Hubble law
even for nearby galaxies ($r<10$ Mpc/h). This raises the question of
why the linear Hubble law is linear at scales where the visible matter
is distributed in-homogeneously.  Several solution to this apparent
paradox have been proposed \cite{wiltshire,baryshev,pwa}: this
situation shows that already the small scale properties of galaxy
distribution have a lot to say on the theoretical interpretation of
their properties.  Indeed, while observations of galaxy structures
have given an impulse to the search for more general solution of
Einstein's equations than the Friedmann one, it is now a fascinating
question whether such a more general framework may provide a different
explanation to the various effects that, within the standard FRW
model, have been {\it interpreted} as Dark Energy and Dark Matter.


\subsection*{Acknowledgements}
I am grateful to T. Antal, Y. Baryshev, A. Gabrielli, M. Joyce, 
M. L\'opez-Corredoira and N. L. Vasilyev for fruitful collaborations,
comments and discussions. I acknowledge the use of the Sloan Digital
Sky Survey data ({\tt http://www.sdss.org}), of the 2dFGRS data ({\tt
  http://www.mso.anu.edu.au/2dFGRS/})of the NYU Value-Added Galaxy
Catalogue ({\tt http://ssds.physics.nyu.edu/}), of the Millennium run
semi-analytic galaxy catalogue ({\tt
  http://www.mpa-garching.mpg.de/galform/agnpaper/})

\end{document}